# Wave localization in stratified square-cell lattices. The antiplane problem

G. G. Osharovich[a], M. V. Ayzenberg-Stepanenko[b]

[a]Department of Mathematics, Bar-Ilan University, Israel
[b]Department of Mathematics, Ben-Gurion University of the Negev, Beer-Sheva, Israel

**Abstract**

Steady-state and transient antiplane dynamic processes in a structured solids consisting of uniform periodic square-cell lattices connected by a lattice layer of different bond stiffnesses and point masses are analyzed. A semi-infinite lattice covered by a layer is also considered. Localization phenomena that are characterized by a waveguide-like propagation along the layer direction and exponential attenuation along its normal are studied. Waveguide pass-bands and attenuation factors are obtained analytically, while transient processes developed under the action of a monochromatic local source are numerically simulated. As a result, it is shown how a two-dimensional problem is transformed with time into a quasi-one-dimensional one and how a layer traps the source energy. Special attention is paid to revealing particularities of transient waves in cases where steady-state solutions are absent: resonant waves with frequencies demarcating pass- and stop-bands at the ends of the Brillouin zone and wave transition in the vicinities of transition points in dispersion curves. In the latter case, a simultaneous onset of different localization phenomena – a spatial star-like beaming and a one-dimensional waveguide-like localization – is shown.

## 1. Introduction

In wave dynamics, there are different areas and scales, which are described by periodic-like models: from atomic lattices, nanofiber networks, composite materials to periodically reinforced casings of aircrafts, earthquake waves in structured massifs and catastrophic waves in power transmission lines. Nowadays, for example, with rapidly progressing nanotechnologies, it is necessary to obtain predictable properties of nanostructures using critical mechanical actions.

Mathematical models and approaches to the analysis of wave propagation in structured materials have a long history. Certainly, great attention was attracted to the periodic mass-spring lattice (MSL) due to its simplicity and physical clearness. One-dimensional models of periodic systems were initially developed in the pioneering work of Lord Rayleigh [1]. The next work that appeared about 70 years later was the fundamental monograph of Brillouin [2] who formulated mathematical aspects of the wave filter using the Floquet theorem for differential equations with periodic coefficients. Later, steady- and unsteady-state wave propagation processes in continuous and periodic-type structures were studied in many works (see, e.g., [3-11]) where various wave localization phenomena were discussed.

It is common knowledge that the *frequency-band localization phenomenon* exists in infinite uniform periodic structures: free sinusoidal wave propagation occurs only in certain discrete bands of frequencies known also as pass-bands. These bands are alternated with stop-bands in which there is no propagation, but spatial decay of sinusoidal waves takes place. This phenomenon has regained new interest since the beginning of the 90s (see, e.g. [12-16]), when artificial crystals were revealed as band-gap materials allowing the control of the propagation of different waves: electronic, electromagnetic, acoustic and vibrational waves.

A *resonant localization phenomenon* in 1D band-gap materials was discovered in [17]. This localization is realized at frequencies in the ends of the Brillouin zone (BZ) demarcating pass and stop bands. In such points, the group velocity $c_g = \partial \omega / \partial k = 0$ ($\omega$ is the frequency, $k$ is the wavenumber) and the energy flow from the source to the periphery is not a wave process, but it is similar to heat propagation (more precisely, the corresponding law of energy propagation depends on the order of the first nonzero derivative $\partial^n \omega / \partial k^n$ at the resonant point). Resonances in quasi-one-dimensional rectangular-cell lattices were recently analyzed in [18]. Recently discovered *beaming phenomena* [19] that are realized in uniform lattices at frequencies within the pass-band are also of a resonant nature. Later diverse star-like beaming patterns were discovered in [20] for a triangular-cell lattice and orthotropic square-cell lattices of massless and material bonds. (Note that we have listed localization phenomena inherent only to regular structures, the well-known localization effects in disordered and random lattices that are not considered below).

In the present work, we study a new type of wave localization in 2D stratified band-gap structures – the waveguide-like one-dimensional propagation of sinusoidal waves. To the authors' knowledge, the first indication of such a localization was manifested in [21] while studying localized transition waves of a mode III fracture in a square-cell lattice with an interface bistable bond layer obeying the constitutive law with two branches. The mentioned waveguide localization phenomena inherent to stratified discrete media are caused by the discreteness and periodicity of the structure components and have no analogues in continuous solids. In [21] it was shown that if a branch determines the bond stiffness higher than that in the surrounding lattice, then sinusoidal waves can be localized along the layer direction and exponentially attenuated along its normal. Afterwards problems of the dynamic crack growth in such a lattice caused by a localized sinusoidal wave were considered in [22-23], while in [24] some related examples were presented of such a localization in layered lattices.

While dispersion patterns in diverse periodic structures were analyzed rather comprehensively, unsteady responses to various kinds of nonstationary excitations were studied to a significantly lesser extent. The latter is notably significant for revealing wave phenomena when the steady state solution is absent, as, for example, in the case of resonant excitation or in the case of the transition from one steady-state regime to another. Probably, the first analytical transient solution for impact propagation in a mass-spring chain (MSC) was presented by Slepyan in the introduction to his monograph [4]. Subsequently, a set of nonstationary dynamic problems was analyzed by united numerical and analytical approaches allowing steady- and unsteady-state processes to be simultaneously studied (see, e.g., [8-10, 18] related to 1D problems and [20-21, 24-30] – to 2D problems).

Following the models and approaches used in [20, 21, 24], we study dynamic antiplane problems for mass-spring bond square-cell structures of two uniform lattices with the interface of a lattice layer of different bond stiffnesses and point masses. A particular case – the semi-infinite lattice covered by a layer – is also considered. Pass-bands of the waveguide and attenuation factors are analytically obtained, while transient processes developing under the action of a monochromatic local source are numerically simulated. We show below how a two-dimensional problem is transformed with time into a quasi-one-dimensional problem and how a layer traps the source energy. Special attention is paid to revealing particularities of transient waves in cases where steady state solutions are absent: resonant waves with frequencies demarcating pass- and stop-bands at the ends of the BZ, and wave patterns in the vicinities of transition points in dispersion curves.

The paper is constructed as follows. In Section 2, equations of motion are formulated for a general geometry of a sandwiched lattice, and dispersion relations are obtained. Then partial cases are specified in Sections 3, 4 and 5. The stratum in Section 3 is a MSC with different masses, while in Section 4, the stratum consists of two linked mass-spring chains with bonds of different stiffnesses along two directions. In Section 5, a layer consisting of three chains is considered. We present dispersion analysis of the system allowing frequency bands of localized waves and decay factors to be obtained depending on layer parameters. Transient processes excited by a local sinusoidal source are numerically simulated. We show the possibility of a simultaneous onset of different localization phenomena – the spatial star-like beaming and the waveguide localization. In Section 4, the analytical solution of the steady-state problem considered in [21] and analytical-numerical solutions of resonantly localized waves are also presented.

## 2. Generalized model of a stratified lattice

Below we formulate a generalized model of an anti-plane lattice approximation for a class of stratified square-cell structures and write a governing system intended for obtaining dispersion relations.

Let $m$ and $n$ $(-\infty < m, n < \infty)$ be numbers of lattice nodes or, which is the same – discrete coordinates. If $l$ is the distance between nodes, then continual coordinates $x$ and $y$ are $x = mh, y = nh$. The considered stratified structure consists of two identical semi-infinite homogeneous lattices $1$ and $3$ connected by a stratum $2$ of $N+1$ $x$-chains (see Fig. 1). Each stratum possesses its own physical parameters: masses and stiffnesses.

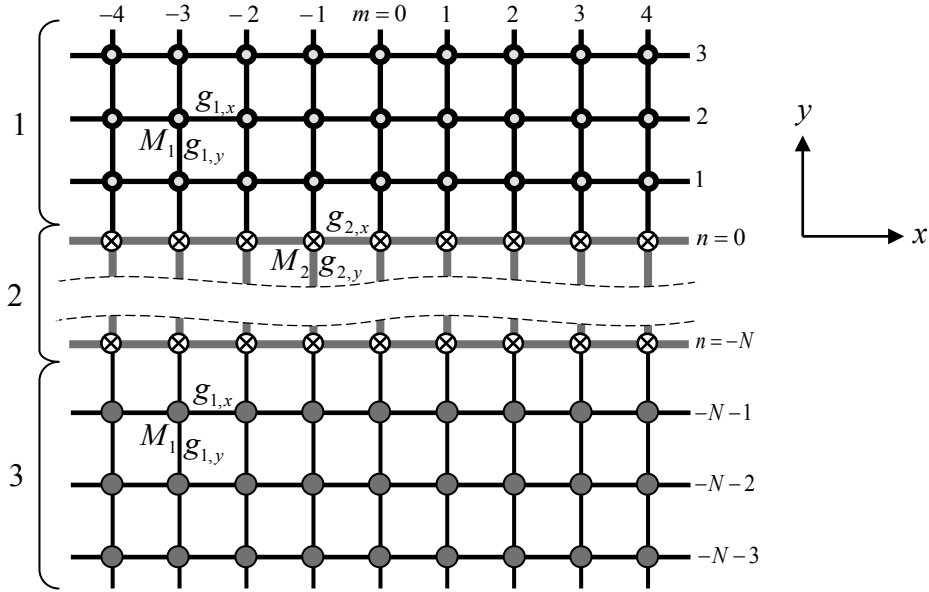

Fig. 1 A square-cell structured lattice with an embedded lattice stratum.

We introduce homogeneous differential equations of the lattice dynamics as follows:

$$\begin{aligned}
M_1 \ddot{u}_{m,n} &= g_{1,x}\left(u_{m+1,n} - 2u_{m,n} + u_{m-1,n}\right) + g_{1,y}\left(u_{m,n+1} - 2u_{m,n} + u_{m,n-1}\right) & ,n > 1 \\
M_1 \ddot{u}_{m,1} &= g_{1,x}\left(u_{m-1,1} - 2u_{m,1} + u_{m+1,1}\right) + g_{2,y}\left(u_{m,0} - u_{m,1}\right) + g_{1,y}\left(u_{m,2} - u_{m,1}\right) & ,n = 1 \\
M_2 \ddot{u}_{m,0} &= g_{2,x}\left(u_{m-1,0} - 2u_{m,0} + u_{m+1,0}\right) + g_{2,y}\left(u_{m,-1} - u_{m,0}\right) + g_{1,y}\left(u_{m,1} - u_{m,0}\right) & ,n = 0 \\
M_2 \ddot{u}_{m,n} &= g_{2,x}\left(u_{m+1,n} - 2u_{m,n} + u_{m-1,n}\right) + g_{2,y}\left(u_{m,n+1} - 2u_{m,n} + u_{m,n-1}\right) & ,-N < n < 0 \quad (1) \\
M_2 \ddot{u}_{m,-N} &= g_{2,x}\left(u_{m-1,-N} - 2u_{m,-N} + u_{m+1,-N}\right) + g_{2,y}\left(u_{m,-N+1} - u_{m,-N}\right) + g_{1,y}\left(u_{m,-N-1} - u_{m,-N}\right) & ,n = -N \\
M_1 \ddot{u}_{m,-N-1} &= g_{1,x}\left(u_{m-1,-N-1} - 2u_{m,-N-1} + u_{m+1,-N-1}\right) + g_{2,y}\left(u_{m,-N} - u_{m,-N-1}\right) + g_{1,y}\left(u_{m,-N-2} - u_{m,-N-1}\right) & ,n = -N-1 \\
M_1 \ddot{u}_{m,n} &= g_{1,x}\left(u_{m+1,n} - 2u_{m,n} + u_{m-1,n}\right) + g_{1,y}\left(u_{m,n+1} - 2u_{m,n} + u_{m,n-1}\right) & ,n < -N-1
\end{aligned}$$

Here $u_{m,n}$ is a transverse displacement of $(m, n)$ node, stiffnesses of x- and y-bonds are denoted by $g_{j,x}$ and $g_{j,y}$ ($j = 1,2$), respectively; particle-in-node masses are denoted by $M_j$. Cell size, $h$, is assumed to be the length unit. Mass and stiffness parameters of the semi-infinite strata *1* and *3* are measurement units - $M_1 = 1$ and $g_{1,x} = g_{1,y} = 1$. As stated above, this lattice structure is intended for describing anti-plane deformation; however, a hypothetic plane model with a single displacement (in x- or y- direction) can also be assumed.

To reveal the existence of the waveguide propagation in the strata direction, we seek a solution to this system in the form of a plane sinusoidal wave propagating in *x*-direction with the magnitude decaying at $n \to \pm\infty$ (for this purpose, factors $\xi_+$ and $\xi_-$ are introduced):

$$\begin{cases} u_{m,n}(t) = \xi_+^n \cdot e^{i(\omega t \pm km)} & , n > 0 \\ u_{m,n}(t) = A_n \cdot e^{i(\omega t \pm km)} & , -N \leq n \leq 0 \\ u_{m,n}(t) = \xi_-^{-(n+N)} \cdot e^{i(\omega t \pm km)} & , n < -N \end{cases} \quad (2)$$

where $\omega$ is frequency and $k$ is the wave number. After substituting (2) into (1), we obtain a system of transcendental equations connecting $\omega$, $k$, factors $\xi_+$, $\xi_-$ and amplitudeparameters $A_n$:

$$\begin{cases} \omega^2 = 4\sin^2(k/2) + (2 - \xi_+ - \xi_+^{-1}) & , n > 1 \\ \omega^2 = 4\sin^2(k/2) + (1 - A_0/\xi_+) + (1 - \xi_+) & , n = 1 \\ M_2\omega^2 = 4g_{2,x}\sin^2(k/2) + g_{2,y}(1 - A_{-1}/A_0) + (1 - \xi_+/A_0) & , n = 0 \\ M_2\omega^2 = 4g_{2,x}\sin^2(k/2) + g_{2,y}(A_{n+1} - 2A_n + A_{n-1})/A_n & , -N < n < 0 \\ M_2\omega^2 = 4g_{2,x}\sin^2(k/2) + g_{2,y}(1 - A_{-N+1}/A_{-N}) + (1 - \xi_-/A_{-N}) & , n = -N \\ \omega^2 = 4\sin^2(k/2) + (1 - A_{-N}/\xi_-) + (1 - \xi_-) & , n = -N - 1 \\ \omega^2 = 4\sin^2(k/2) + (2 - \xi_- - \xi_-^{-1}) & , n < -N - 1. \end{cases} \quad (3)$$

In some simple cases (see below), the dispersion relation $\omega = \omega(k,...)$ can be analytically obtained, while for others it is numerically calculated.

The system (3) comprises at most $N + 5$ equations possessing known mass and stiffness parameters and $N + 5$ variables: $\xi_+$, $\xi_-$, $A_n$, $k$ and $\omega$. Such a nonlinear system has obviously no analytical solution in the common case. Below some partial systems are considered, for which dispersion relations are analytically obtained. We have found and analyzed only dispersion relations satisfying conditions $|\xi_+| < 1$ and $|\xi_-| < 1$ determining the waveguide pattern of propagation wave. Inequalities $0 < \xi < 1$ and $-1 < \xi < 0$ result in in-phase and anti-phase oscillations of neighboring nodes in the $n$- direction, respectively. (Note that the attenuation of wave amplitudes in the $n$-direction is inversely proportional to $|\xi|$, but for convenience, below we call $\xi$ the decay factor).

## 3. Lattice with an embedded chain of masses
### 3.1. Dispersion analysis
#### 3.1.1. *Infinite lattice*

Consider a single chain with node mass $M_2 \equiv M$ embedded inside a homogeneous lattice (as shown in Fig. 2). Node masses and bond stiffnesses of the latter are taken as measurement units.

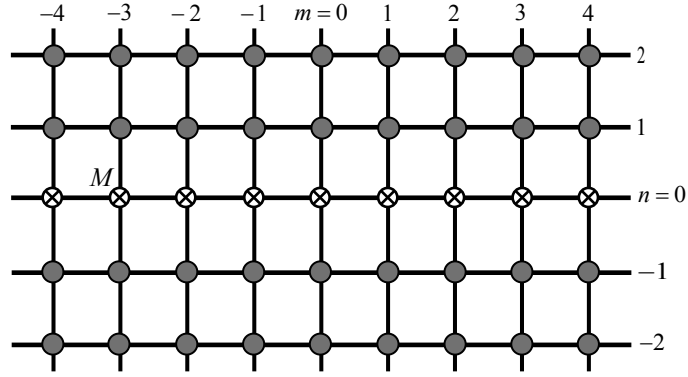

Fig. 2 A chain stratum embedded into a lattice.

In this case, free wave propagation in the structure is described by the following system:

$$\begin{cases} \ddot{u}_{m,n} = \left(u_{m+1,n} + u_{m-1,n} + u_{m,n+1} + u_{m,n-1} - 4u_{m,n}\right) & , n < 0 \\ M\ddot{u}_{m,0} = \left(u_{m-1,0} - 2u_{m,0} + u_{m+1,0}\right) + \left(u_{m,1} - u_{m,0}\right) + \left(u_{m,-1} - u_{m,0}\right) & , n = 0 \\ \ddot{u}_{m,n} = \left(u_{m+1,n} + u_{m-1,n} + u_{m,n+1} + u_{m,n-1} - 4u_{m,n}\right) & , n > 0. \end{cases} \quad (4)$$

Our aim is to detect the waveguide propagation of elastic waves along the interface stratum. We seek a solution to the system (4) in the form of a plane moving wave with the magnitude vanishing at $n \to \pm\infty$. Due to the symmetry of (4), the attenuation factors $\xi_+ = \xi_- \equiv \xi$, $|\xi| < 1$.

$$\begin{cases} u_{m,n}(t) = \xi^{-n} \cdot e^{i(\omega t \pm km)} & , n \leq -1 \\ u_{m,n}(t) = A_0 e^{i(\omega t \pm km)} & , n = 0 \\ u_{m,n}(t) = \xi^{n} \cdot e^{i(\omega t \pm km)} & , n \geq 1. \end{cases} \quad (5)$$

Substituting (5) into (4) we obtain:

$$\begin{cases} \omega^2 = 4\sin^2(k/2) + 2 - \xi - \xi^{-1} \\ \omega^2 = 4\sin^2(k/2) + 2 - A_0/\xi - \xi \\ M\omega^2 = 4\sin^2(k/2) + 2(1 - \xi/A_0). \end{cases} \quad (6)$$

Then we have detected the existence of real dependences $\omega = \omega(k)$ taking into account the condition $|\xi| < 1$ and thereby proved the waveguide propagation.

The dispersion relation for the frequency is the first equation (6) rewritten as

$$\omega = \sqrt{2 + \Psi - \Gamma} \quad (\Psi = 4\sin^2(k/2),\ \Gamma = \xi + \xi^{-1}). \quad (7)$$

The two first equations of system (6) result in $A_0 = 1$; from the second and the third ones we have obtained the following square equation for determining $\xi$:

$$(2 - M)\xi^2 - (1 - M)(\Psi + 2)\xi - M = 0. \quad (8)$$

Roots $\xi(M,k)$ of (8) are real for any $M$ and monotonic in the intervals $0 < M < 1$, $1 < M < \infty$. In these intervals only one of two roots satisfies the condition $|\xi| < 1$. Note that in case of $M = 2$, Eqn. (8) possesses a single root – $\xi = M(\Psi + 2)^{-1}$. Thus, the waveguide propagation exists in the considered system for any $M$ except $M = 1$, where $\xi = \pm 1$; in the latter case, trivial 1D oscillation forms along $n$-direction occur with constant in-phase ($\xi = 1$) and anti-phase ($\xi = -1$) amplitudes. The condition $|\xi| < 1$ is satisfied for all $k$ within the BZ ($0 \leq k \leq \pi$), the sign of $\xi$ depending on $M$ in the following manner:

$$\xi(M) = \begin{cases} 0 < \xi < 1, & M > 1 \\ -1 < \xi < 0, & 0 < M < 1. \end{cases} \quad (9)$$

The anti-phase (in-phase) oscillation form is realized along the $n$-direction if $M > 1$ ($M < 1$).

In Fig. 3 dependencies $\xi(k)$, $\omega(k)$ and $c_g(k) = \dfrac{\partial \omega}{\partial k}$ (recall that $c_g$ is the group velocity) are depicted for various values of $M$. Dashed curves correspond to $M < 1$, while solid curves to $M > 1$. One can see that the frequency, group velocity and pass-band width $\Delta\omega = \omega(\pi) - \omega(0)$ decrease with increasing $M$. The case $M = 1$ is presented by dash-dotted and dotted curves for two limiting values: $\xi = +1$ and $\xi = -1$, respsectively.

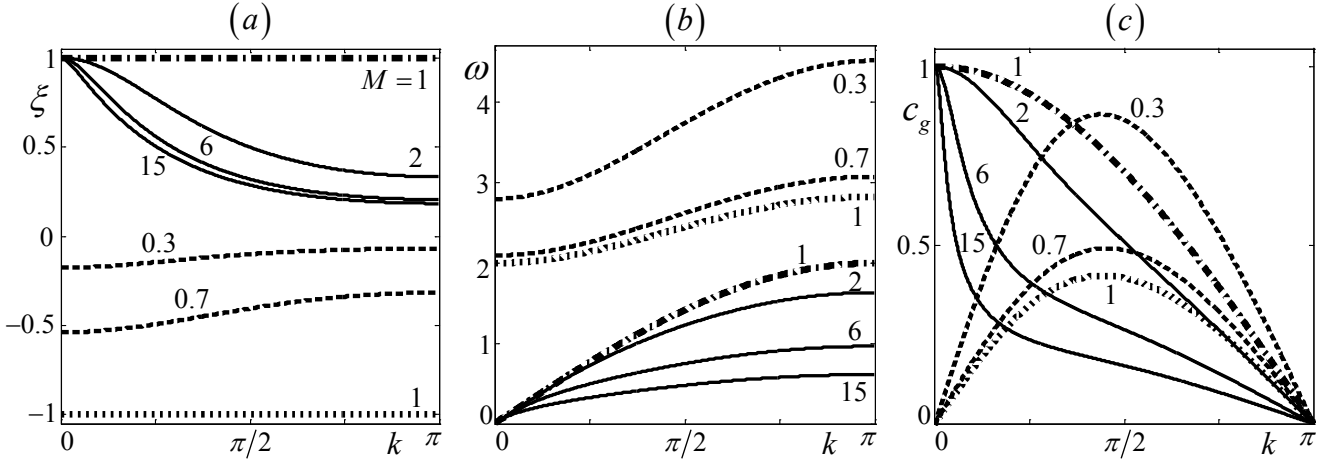

Fig. 3 Decay factor $\xi(k)$ - (a) and dispersion curves, $\omega(k)$ - (b) and $c_g(k)$ - (c) for several values of $M$.

Frequencies and group velocities in this case are:

$$\begin{aligned}\xi = 1 &: \omega = 2\sin(k/2),\ c_g = \cos(k/2); \\ \xi = -1 &: \omega = 2\sqrt{1+\sin^2(k/2)},\ c_g = \sin(k)/\omega.\end{aligned} \quad (10)$$

The first dispersion relation (10) coincides with that of the MSC ($M = g = 1$) − the dash-dotted curve in Fig. 3(b), while the second one is the dispersion relation for the MSC on an elastic foundation of stiffness $g_* = 1$ − the dotted curve in Fig. 3(b).

The similarities and differences of dispersion patterns inherent to the 2D in-phase oscillations for $M = 1.05, 4$, and $30$ ($0 < \xi < 1$) and to the MSC with the same values of $M$ can be detected in Fig. 4 by comparison of solid and dashed curves, respectively. The anti-phase case ($0 < M < 1$, $-1 < \xi < 0$) is particularly discussed below, in Section 4.

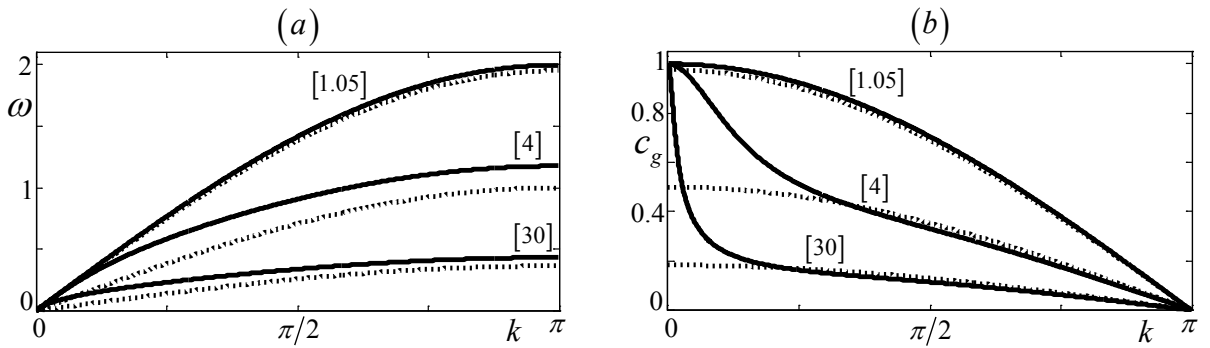

Fig. 4 Comparison of dispersion curves for a stratified lattice (a) and a MSC (b).

If $M$ is close to unity, a good correspondence of comparable curves is obvious and does not require a special explanation. With increasing $M$, their relative difference in the longwave spectrum ($k$ is small) can be significant, while if $k$ increases, dispersion patterns converge (this is particularly noticeable for group velocities: in the presented examples, the difference between group velocities is not practically detectable for $k > 1.139$ ($M = 4$) and $k > 0.668$ ($M = 30$)). Such a behavior complies with the following explanation. If $k$ is small, the decay factor $\xi$ is close to unity – Fig. 3 (*a*). This determines a weak attenuation of oscillations along *y*-direction. Then $n = 0$ chain and a (thick) surrounding lattice layer oscillates uniformly – with practically the same amplitudes (the thickness of the layer tends to infinity with $k \to 0$). To this end, a relative average chain mass within the layer tends to $1^+$ even if the chain $n = 0$ has a relatively large mass $M$. Therefore, the lattice dispersion, which is similar to that in the MSC (with $M = 1$), turns out to be detectably different in comparison with the MSC of a corresponding $M$.

If $k$ increases, the decay factor $\xi$ decreases (or, which is the same, –the attenuation in *y*-direction increases), and the thickness of the above-mentioned "uniform" layer decreases rapidly. Thus, the contribution of a "heavy" chain rises and becomes decisive for relatively large values of $k$. Since $\xi$ also decreases with increasing $M$, the part of the longwave spectrum characterized by a significant difference is shorter for larger $M$.

### 3.1.2. *Semi-infinite lattice bounded by a chain*

Consider a semi-infinite lattice (let it be the upper half-plane) of parameters $M_1 = g_{1,x} = g_{1,y} = 1$ bounded by a chain with the node mass $M$ and spring stiffness $g_{2,x} = 1$. For this structure, the common case (3) results in the following system:

$$\begin{cases} M\ddot{u}_{m,0} = u_{m-1,0} - 2u_{m,0} + u_{m+1,0} + \left(u_{m,1} - u_{m,0}\right) &, n = 0 \\ \ddot{u}_{m,n} = u_{m+1,n} + u_{m-1,n} + u_{m,n+1} + u_{m,n-1} - 4u_{m,n} &, n > 0. \end{cases} \qquad (11)$$

Substituting the "waveguide" solution

$$\begin{cases} u_{m,n}(t) = A e^{i(\omega t \pm km)} &, n = 0 \\ u_{m,n}(t) = \xi^n \cdot e^{i(\omega t \pm km)} &, n \leq 1 \end{cases} \qquad (12)$$

into (11), we obtain:

$$\begin{cases} \omega^2 = 4\sin^2(k/2) + 2 - \xi - \xi^{-1} \\ \omega^2 = 4\sin^2(k/2) + 2 - A_0/\xi - \xi \\ M\omega^2 = 4\sin^2(k/2) + (1 - \xi/A_0). \end{cases} \quad (13)$$

In the considered case, we have obtained $A_0 = 1$, and the dispersion relation has the same expression (7) as in the previous case:

$$\omega = \sqrt{2 + \Psi - \Gamma}, \quad (14)$$

while $\xi$ is determined as a solution of the square equation:

$$(1-M)\xi^2 + \left((M-1)\Psi + 2M - 1\right)\xi - M = 0 \quad (15)$$

which differs from (8). Eqn. (15) with the requirement $|\xi| < 1$ results in the following system of inequalities:

$$\begin{cases} -1 < \xi(k,M) < 0, & 0 < M \le 0.5, \quad k \in [0,\pi] \\ -1 < \xi(k,M) < 0, & 0.5 < M \le 0.75, \quad k \in [k_*,\pi], \quad k_* = 2\cdot\arcsin\left(\sqrt{\dfrac{1-2M}{2M-2}}\right). \\ 0 < \xi(k,M) < 1, & 1 < M, \quad k \in [0,\pi] \end{cases} \quad (16)$$

In the case of $M = 1$, we obtain a trivial solution: $\xi = 1$, oscillations along the $n$-direction are in-phase and possess the same amplitude as in the corresponding MSC.

The main difference between infinite and semi-infinite lattices is the surprising behavior of the dispersion in the latter. There exists an interval $0.5 < M \le 0.75$, in which the requirement $|\xi| < 1$ is satisfied only in the part of the BZ, $[k_*(M), \pi]$, not including long wavelengths. The localization is realized if $k > k_*$, which are called below *limiting wave numbers*, and there is no localization otherwise. The dependence $k_*(M)$ is monotonously increasing from $k_*(0.5) = 0$ to $k_*(0.75) = \pi$. Moreover, within the interval $0.75 < M \le 1$, Eqn. (15) has no waveguide solutions determining $|\xi| < 1$ in the entire BZ. Such a behavior contradicts the infinite lattice, where the requirement $|\xi| < 1$ is satisfied within the entire BZ. Below, frequencies corresponding to limiting wave numbers points, $\omega_*$, are called "transition points".

In Fig. 5, dependencies $\xi(k)$, $\omega(k)$ and $c_g(k)$ are depicted for various values of $M$. The curves for $M > 1$ are plotted by solid lines, for $0 < M \le 0.5$ — by dotted lines, and for $0.5 < M \leq 0.75$ – by bold dashed lines. As one can see, differences in dispersion patterns of infinite and semi-infinite lattices practically vanished in the case $M > 1$ (compare with corresponding curves in Fig. 3) determining in-phase oscillations along the $n$-direction.

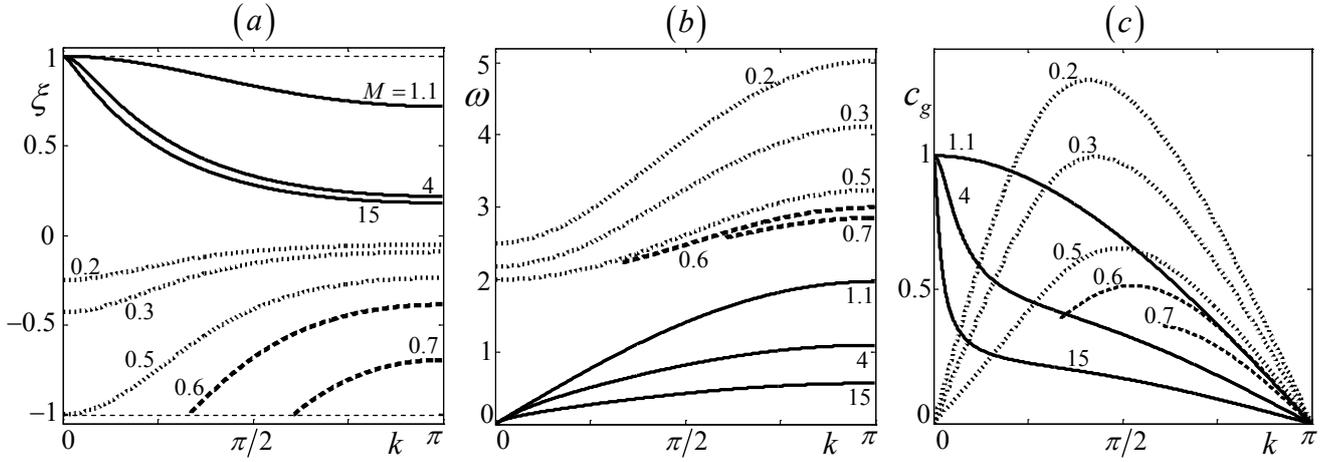

Fig. 5 Dispersion curves $\xi(k)$ - (a), $\omega(k)$ - (b) and $c_g(k)$ - (c) for a semi-infinite lattice.

Such unexpected peculiarities of waves with parameters located in the vicinities of transition points can be revealed by analyzing the transient problem (note that the steady-state solution does not exist for the transition frequency $\omega_*$). Related results are presented below, in Subsection 3.2.2.

### 3.2. Transient solution

We have numerically obtained transient solutions of the above-considered problems in the case of the action of kinematic sinusoidal source located in the $(0,0)$ node:

$$u_{0,0} = \sin(\omega_0 t) H(t) \qquad (17)$$

where $\omega_0$ is the source frequency and $H(t)$ is the Heaviside step function. Condition (17) and zero initial conditions are added to equations of motions (4) and (11).

Computer simulations of these problems were conducted using explicit finite-difference algorithms. Calculations show that the unsteady-state response, displacements $u_{m,n}(t)$, can be accurately represented in the form $u_{m,n}(t) = U_{m,n}(t)\sin(\omega_0 t)$, where $U_{m,n}(t)$ is the envelope of perturbations with the carrier frequency $\omega_0$. Numerical results are shown below as dependences $U_{m,n}(t)$ for diverse $m$ and $n$ and as snapshots of $|U_{m,n}|$ at fixed values of $t$.

### 3.2.1. *Infinite lattice*

Curves in Fig. 6 are snapshots of $|U_{m,n}|$ at $m \geq 0$ (below – "envelopes") for chains $n = 0, 1, 2$ and 5 at $t = 500$. Three examples of calculations: $(a)\, M = 2$, $(b)\, M = 6$ and $(c)\, M = 15$ are shown. Source frequency $\omega_0$ corresponds to the middle of the pass-band for each $M$. The chosen $M$ and $\omega_0$ determine corresponding dispersion parameters: wave number, $k_0$, decay factor, $\xi$, and group velocity, $c_0$ (see Table 1).

Table 1 Dispersion parameters for various $M$

| $M$ | $\omega_0$ | $k_0$  | $\xi$  | $c_0$  |
|-----|-----------|--------|--------|--------|
| 2   | 0.8165    | 0.9115 | 0.7208 | 0.7355 |
| 6   | 0.4826    | 0.7576 | 0.5750 | 0.4051 |
| 15  | 0.3064    | 0.7129 | 0.5395 | 0.2457 |

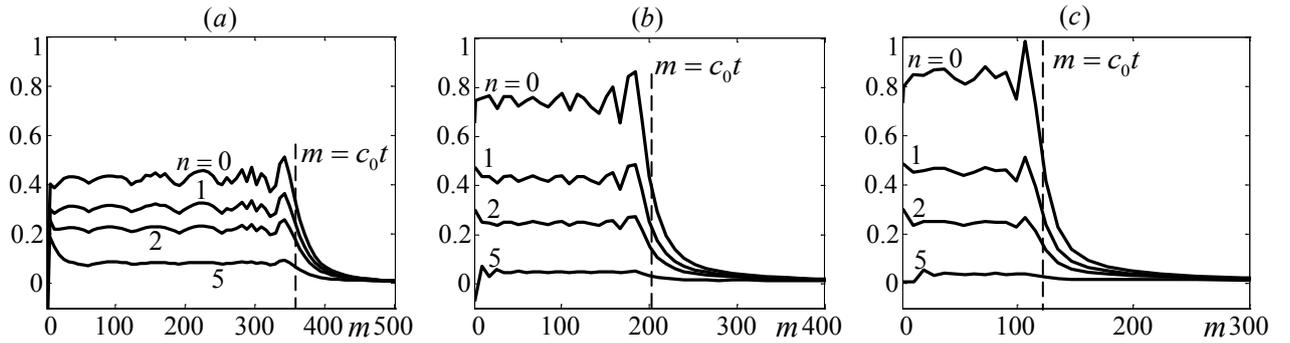

Fig. 6 Envelopes $|U_{m,n}|$ versus $m$ for $n = 0, 1, 2$ and 5 at $t = 500$: $(a)\, M = 2$, $(b)\, M = 6$, $(c)\, M = 15$.

In accordance with $\xi(M)$ dependences – see Fig. 3 $(a)$, the greater $M$ (or, what is the same - the lower $\xi$), the stronger the localization effect. A good correspondence can be seen between the propagation velocity of the wave with detectable amplitudes and the group velocity predicted by the dispersion analysis. The maximal contribution of the nonstationary part, relatively small in itself, is observed in the vicinity of the source and in the extending (with increasing $M$) quasi-front area moving with the velocity $c_0$ (vertical dashed lines). The envelopes versus time are presented in Fig. 7: curves numbered 0, 1, 2 and 5 are envelopes versus displacement time in the nodes $(100, 0)$, $(100, 1)$, $(100, 2)$ and $(100, 5)$. The dashed horizontal lines show numerically obtained steady-state solutions, the dotted vertical line shows the moment of the quasi-front arrival into nodes with $m = 100$.

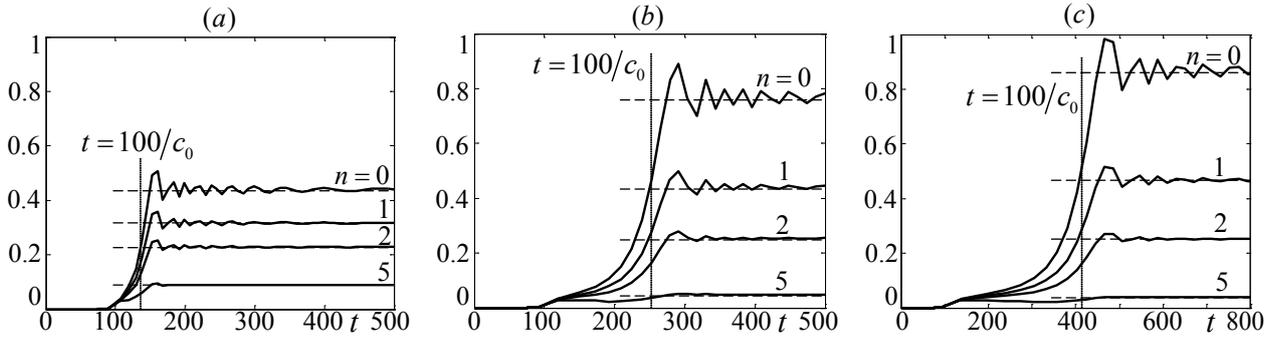

Fig. 7 Envelopes $|U_{100,n}|$ for $n = 0$, 1, 2 and 5 as functions of $t$. (*a*) $M = 2$, (*b*) $M = 6$, (*c*) $M = 15$.

The above results are related to cases where nonstationary solutions have stationary limits. The steady-state amplitudes are reached very quickly after the arrival of perturbations to an examined node (chain); distributions of magnitudes in *n*-chains are in a good accordance with appropriate decay factors. The main results illustrated by these examples are steady-state solutions obtained numerically and estimation of the contribution of time-dependent part. In the following subsection, an analysis of wave patterns is made in the case where frequency $\omega_0$ is located in the vicinity of the transition point $\omega_*$ inside the BZ, when the steady-state limit does not exist or a very large time is needed to reveal it.

3.2.2. *Semi-infinite lattice*

Below we discuss peculiarities of wave patterns corresponding to semi-infinite lattices bounded by a chain in the case where dispersion curves possess transient points inside the BZ. The depicted contours in Fig. 8, (*a*) and (*b*) ($M = 0.6$ and 0.7, respectively) are taken at $t = 200$ snapshots of domain boundaries, beyond which displacements remain less than 10% (solid curves) and 15% (dotted curves) of the maximal amplitude in the source. Such percentages were chosen with the aim to reveal a localized character of the wave pattern. The source frequencies are located in the vicinities of transition points $\omega_*$. In case of $M = 0.6$ ($k_* = 1.0472$, $\omega_* = 2.2361$), the chosen exciting frequencies are $\omega_0 = 2.2$, 2.27 and 2.3; in case of $M = 0.7$ ($k_* = 1.9106$, $\omega_* = 2.5820$), $\omega_0 = 2.5$, 2.65 and 2.7. Some detectable differences in amplitude distribution patterns in these two cases are probably explained by dispersion features: the latter case has noticeably smaller values of "waveguide" parameters – the wavelength zone of the localization, the attenuation level, the passband width and the group velocity (see Fig. 5).

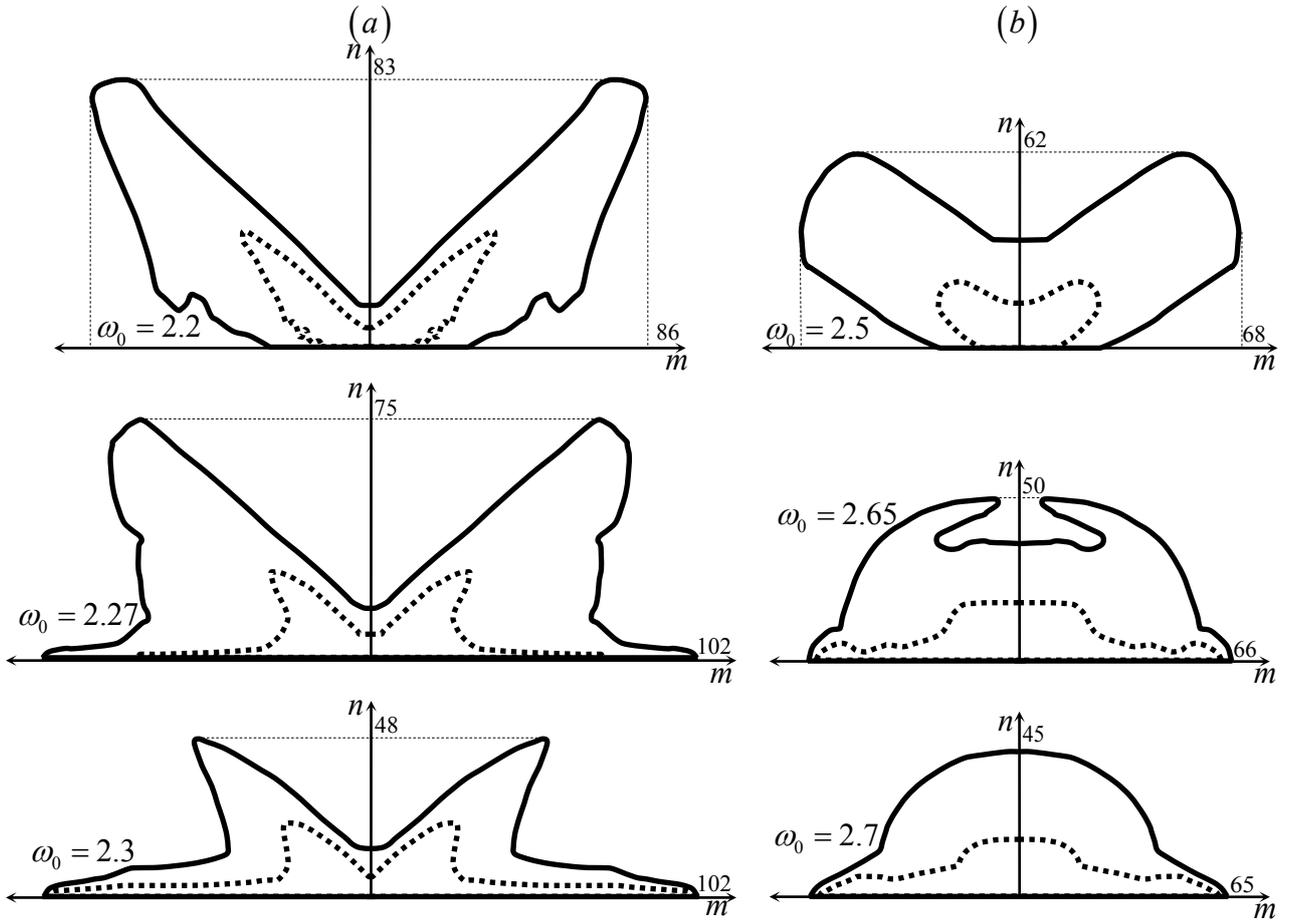

Fig. 8 Contours demarcating domains of nodes oscillating with different amplitudes at $t = 200$. Outside the solid (dotted) contour, absolute values of displacements remain less than 10% (15%) of the maximal value in the source. (*a*) - $M = 0.6$, (*b*) - $M = 0.7$.

Frequencies $\omega_0 = 2.2$ (left column) and $\omega_0 = 2.5$ (right column) are located outside pass bands (see Fig. 5). In these cases, the waveguide localization is absent. The presentation of these results here is motivated by the desire to identify the wave pattern in the vicinity of the transition point and then to indicate the initiation of the steady-state process. Note that spatial distributions of waves at these frequencies have turned out to be qualitatively similar to beaming patterns in a homogenous square-cell lattice possessing prevailing diagonal directions of waves propagation [20]. The closer the frequency $\omega_0$ to the transition point, the faster the beaming area decrease. The initiation and development of both 1D waveguide propagation and the beaming pattern are realized simultaneously in the region adjacent to the source. The influence of the latter is more detectable in case of $M = 0.7$, where the "waveguide" activity is much weaker. After passing through the transition point and a further increase in the excitation frequency, the waveguide pattern becomes prevalent.

## 4. Two-chain stratum

### 4.1. Dispersion analysis

#### 4.1.1. *Infinite lattice*

Consider a stratified structure consisting of a lattice ($M = g_x = g_y = 1$) with an interfaced stratum of two chains possessing node masses $M_2 = 1$ and stiffnesses $\gamma_n \equiv g_{2,y}$, $\gamma_m \equiv g_{2,x}$ (let the chains be $n = 0, -1$ as shown in Fig. 9).

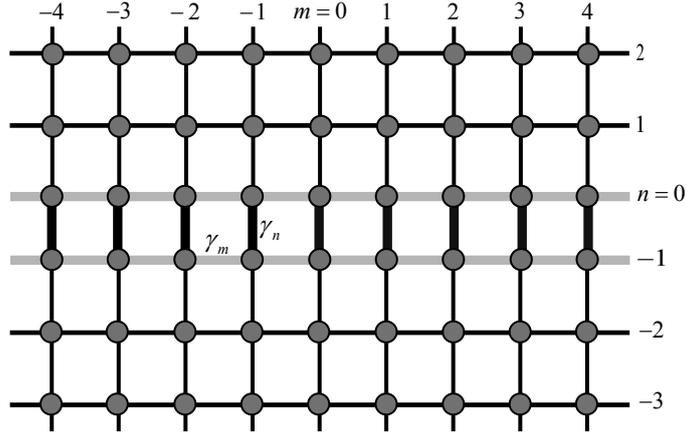

Fig. 9 Lattice with embedded stratum ($n = -1, 0$) with different stiffnesses of bonds.

Free wave propagation in the structure is described by the following system:

$$\begin{cases} \ddot{u}_{m,n} = u_{m+1,n} + u_{m-1,n} + u_{m,n+1} + u_{m,n-1} - 4u_{m,n} & , n \neq 0, -1 \\ \ddot{u}_{m,0} = \gamma_m \left( u_{m-1,0} - 2u_{m,0} + u_{m+1,0} \right) + \left( u_{m,1} - u_{m,0} \right) + \gamma_n \left( u_{m,-1} - u_{m,0} \right) & , n = 0 \\ \ddot{u}_{m,-1} = \gamma_m \left( u_{m-1,-1} - 2u_{m,-1} + u_{m+1,-1} \right) + \left( u_{m,-2} - u_{m,-1} \right) + \gamma_n \left( u_{m,0} - u_{m,-1} \right) & , n = -1. \end{cases} \quad (18)$$

We seek a solution of (18) in the anti-symmetric form:

$$\begin{cases} u_{m,n}(t) = \xi^n \cdot e^{i(\omega t \pm km)} & , n > 0 \\ u_{m,n} = A_n \cdot e^{i(\omega t \pm km)} & , n > 0 \\ u_{m,n}(t) = -\xi^{-(1+n)} \cdot e^{i(\omega t \pm km)} & , n < -1. \end{cases} \quad (19)$$

After substituting (19) into (18) we have found $A_0 = A_{-1} = 1$ and the system of relations:

$$\begin{cases} \omega^2 = 4\sin^2(k/2) - (\xi-1)^2/\xi \\ \omega^2 = 4\gamma_m \sin^2(k/2) + 2\gamma_n - \xi + 1 \end{cases} \quad (20)$$

from which decay factor, $\xi$, and dispersion relation, $\omega = \omega(k, \xi)$, are explicitly obtained as:

$$\xi = \left[\Psi \cdot (1-\gamma_m) - 2\gamma_n + 1\right]^{-1}, \quad \omega = \sqrt{2+\Psi-\Gamma}. \tag{21}$$

We set $\gamma_n \neq 1$ and consider separately two particular cases: (*a*) $\gamma_m = 1$ and (*b*) $\gamma_m \neq 1$.

(*a*) $\gamma_m = 1$. In this case, we arrive at relations obtained in [21]:

$$\xi = \frac{1}{1-2\gamma_n}, \quad \omega = \sqrt{\frac{4\gamma_n^2}{2\gamma_n - 1} + \Psi}. \tag{22}$$

The condition $|\xi| < 1$ is satisfied if $\gamma_n > 1$. In this case $\xi$ is negative, which determines anti-phase oscillations of neighboring *n*-nodes. The limiting case $\gamma_n = 1$ ($\xi = -1$) results in the anti-phase oscillations form in a uniform square MSL.

One can see that frequency $\omega$ in (22) coincides with that in MSC upon an elastic foundation. Homogeneous equations of its motion and the dispersion relation are:

$$\ddot{u}_n = u_{n+1} - 2u_n + u_{n-1} - g_* u_n, \quad \omega = \sqrt{\Psi + g_*}, \tag{23}$$

where $g_*$ is the foundation stiffness. In this case, $g_* = \dfrac{4\gamma^2}{2\gamma - 1}$, and the considered 2D problem can be studied on the basis of the simpler 1D model.

Dependences $\omega(k)$ and $c_g = \sin k / \omega(k)$ for some values of $\gamma$ are depicted in Fig. 10, (*a*) and (*b*), respectively; frequencies and group velocities for $\gamma_n = 1$ are shown by dashed curves and determine lower limiting frequencies and upper limiting group velocities of waveguide propagation.

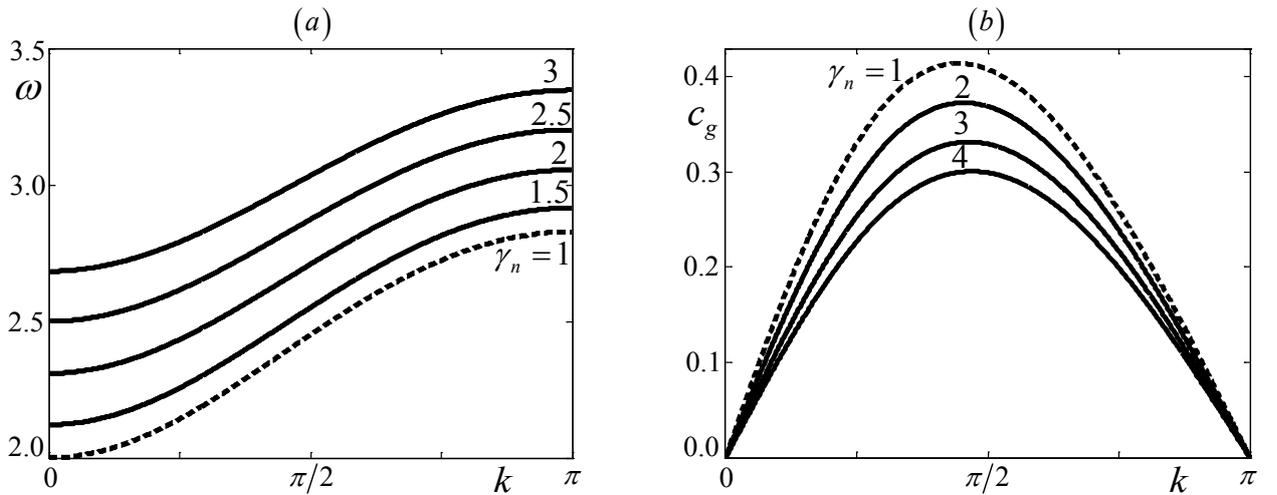

Fig. 10 Dispersion curves for frequency $\omega$ - (*a*) and group velocity $c_g$ - (*b*) at various values of $\gamma_n$.

Frequencies at the ends of BZ ($k = 0$ and $k = \pi$)

$$\omega(0) = 2\sqrt{\frac{\gamma_n^2}{2\gamma_n - 1}}, \quad \omega(\pi) = 2\sqrt{\frac{\gamma_n^2}{2\gamma_n - 1} + 1}, \tag{24}$$

demarcate pass- and stop-bands.

The greater $\gamma_n$, the higher frequencies within a pass-band, while its width, $\Delta\omega = \omega(\pi) - \omega(0)$, as well as $c_g$, decrease with increasing $\gamma_n$. Maxima of group velocities are located near the pass-band median, $k = \pi/2$, where wavelength $\lambda \sim 4$.

(b) $\gamma_m \neq 1$. In this case, we have obtained some different dispersion features depending on the values $\gamma_n$ and $\gamma_m$. First of all, the system of inequalities exists,

$$\begin{cases} \gamma_n + 2\gamma_m > 3 \\ \gamma_n > 1, \ \gamma_m > 0 \end{cases}, \tag{25}$$

which satisfies the requirement $|\xi| < 1$ and thus determines the waveguide solution of system (19) for any $k$ inside the BZ. At the same time, some relations between $\gamma_n$ and $\gamma_m$ result in the waveguide pattern only at a part of the BZ. Such a phenomenon indicated above in Section 3 is much more complicated and varied here. Two kinds of limiting points are detected: in the long wavelength spectrum (the waveguide propagation is proved if $k_* < k \leq \pi$) and in the short wavelength spectrum (with the waveguide propagation if $0 \leq k < k_{**}$). Limiting points, $k_*$, $k_{**}$, corresponding to the condition $|\xi(k)| = 1$ are obtained as:

$$k_* = \begin{cases} 2 \cdot \arcsin\left(\sqrt{\dfrac{\gamma_n}{2(1-\gamma_m)}}\right), & 0 < \gamma_n \leq 1, \ 0 < \gamma_m < 1, \ \gamma_n + 2\gamma_m < 2 \\ 2 \cdot \arcsin\left(\sqrt{\dfrac{(1-\gamma_n)}{2(\gamma_m - 1)}}\right), & 0 < \gamma_n \leq 1, \ \gamma_m > 1, \ \gamma_n + 2\gamma_m > 3 \end{cases} \tag{26}$$

$$k_{**} = 2 \cdot \arcsin\left(\sqrt{(1-\gamma_n)/2(\gamma_m - 1)}\right), \ 0 < \gamma_m < 1, \ \gamma_n > 1, \ 2.5 < \gamma_n + 2\gamma_m < 3.$$

The systems of inequalities:

$$\begin{aligned}
1: & \quad (\gamma_n > 1) \cap (\gamma_m > 0) \cap (\gamma_n + 2\gamma_m > 3) \\
2.1: & \quad (0 < \gamma_n \leq 1) \cap (0 < \gamma_m < 1) \cap (\gamma_n + 2\gamma_m < 2) \\
2.2: & \quad (0 < \gamma_n \leq 1) \cap (\gamma_m > 1) \cap (\gamma_n + 2\gamma_m > 3) \\
3: & \quad (0 < \gamma_m < 1) \cap (2.5 < \gamma_n + 2\gamma_m < 3)
\end{aligned} \tag{27}$$

form the domain in which localization phenomena take place (see Fig. 11). Each inequality of (27) determines the corresponded sub-domain marked by numbers 1, 2.1, 2.2 and 3, respectively. Calculations show that in the white sub-domain in Fig. 11, where none of inequalities (27) is satisfied, there are discrete pairs $\gamma_m$, $\gamma_n$ that determine $|\xi|<1$ for some values of $k$, and thus the existence of the waveguide localization, but we have no analytical description of these cases.

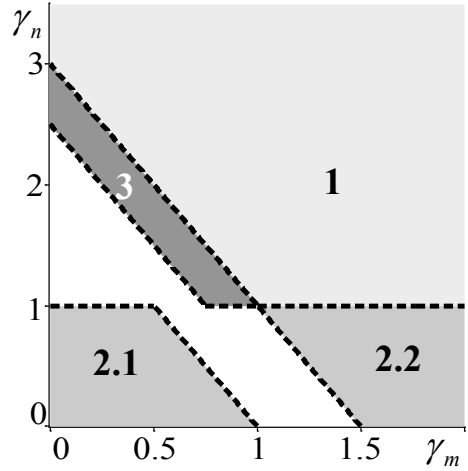

Fig. 11 Sub-domains 1, 2.1, 2.2 and 3 determining the existence of the waveguide localization.

In Fig. 12, dependencies $\xi(k)$ calculated for some pairs of values $\gamma_m$, $\gamma_n$ (shown in square brackets) are depicted. Drawings (*a*), (*b*) and (*c*) correspond to sub-domains 1, 2.1, 2.2 and 3. Dashed parts of the curves correspond to cases in which the condition $|\xi(k)|<1$ is not satisfied and the wave localization is not realized.

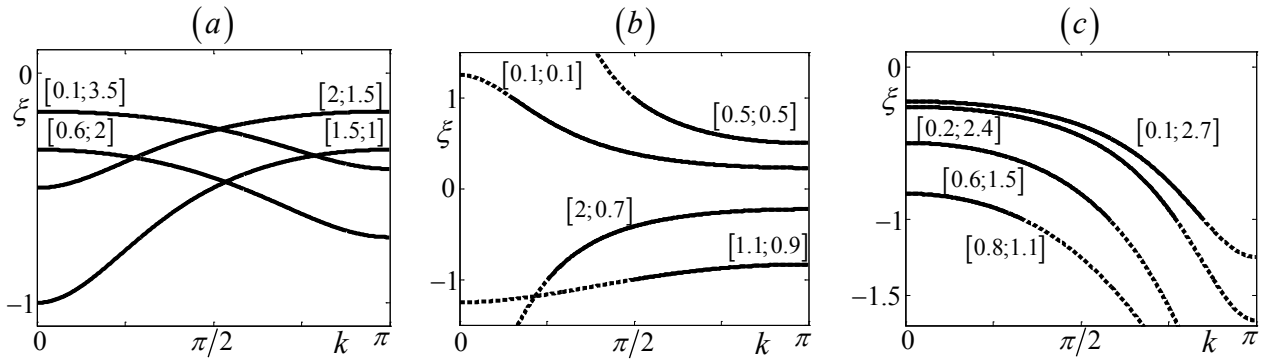

Fig. 12 Decay factor $\xi$ - versus $k$ for various values of $\gamma_m, \gamma_n$ (in square brackets). Figures (*a*), (*b*) and (*c*) correspond to sub-domains 1, 2.1, 2.2 and 3 in Fig. 11.

In Fig. 13, dependencies $\omega(k)$ are depicted for the same pairs of $\gamma_m$ and $\gamma_n$ as in Fig. 12.

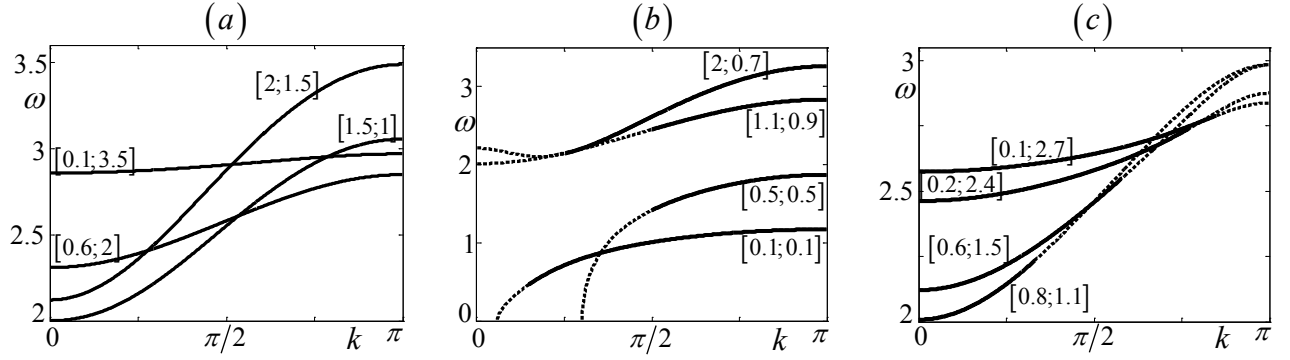

Fig. 13 Dispersion curves $\omega(k)$ for various pairs $[\gamma_m, \gamma_n]$. Figures (*a*), (*b*) and (*c*) correspond to sub-domains 1, 2.1, 2.2 and 3 in Fig. 11.

For dashed parts of the curves, condition $|\xi(k)| < 1$ is not satisfied. The "waveguide" pass-bands are bounded by lower (upper) transition points $\omega_*$ ($\omega_{**}$) corresponding to limiting wave number $k_*$ ($k_{**}$).

As one can see from Eqn. (21) and Fig. 13(*a*), the pass-band, $\Delta\omega = \omega(\pi) - \omega(0)$, strongly depends on $\gamma_m$ (the larger $\gamma_m$, the wider $\Delta\omega$) and weakly – on $\gamma_n$.

### 4.1.2. *Semi-infinite lattice*

If we set $\gamma_n = 0$ in Eqn. (18), we obtain a model of a semi-infinite lattice bounded by a chain with $M = 1$ and stiffness of *x*-bonds $\gamma_m$. Let such a lattice be located in the upper half of the plane and $\gamma_m \neq 1$. Free wave propagation in the structure is described by the following system:

$$\begin{cases} \ddot{u}_{m,n} = u_{m+1,n} + u_{m-1,n} + u_{m,n+1} + u_{m,n-1} - 4u_{m,n} & (n \neq 0) \\ \ddot{u}_{m,0} = \gamma_m \left( u_{m-1,0} - 2u_{m,0} + u_{m+1,0} \right) + \left( u_{m,1} - u_{m,0} \right). \end{cases} \quad (28)$$

As above, we seek the waveguide solution in the form $u_{m,n}(t) = \xi^n \cdot e^{i(\omega t \pm km)}$ with the condition $|\xi| < 1$ taken into account. The dispersion relation and decay factor have the same expression (21) as in the previous case:

$$\omega = \sqrt{2 + \Psi - \Gamma}, \quad \xi = \left[ \Psi \cdot (1 - \gamma_m) + 1 \right]^{-1}. \quad (29)$$

Eqn. (29) with the requirement $|\xi| < 1$ results in the following system of inequalities:

$$\begin{cases} 0 < \xi(k,\gamma_m) < 1, & 0 < \gamma_m < 1, \quad k \in [0,\pi] \\ -1 < \xi(k,\gamma_m) < 0, & 1.5 < \gamma_m, \quad k \in (k_*,\pi], \; k_* = 2\cdot\arcsin\left(\sqrt{\dfrac{1}{2(\gamma_m-1)}}\right). \end{cases} \qquad (30)$$

As in previous cases, there exist transient points $\omega_*$ inside the BZ.

In Fig. 14, relations $\xi(k)$, $\omega(k)$ and $c_g(k)$ are depicted for various values of $\gamma_m$ marked in square brackets. All the curves are depicted only in the range of $k$ where $|\xi| < 1$.

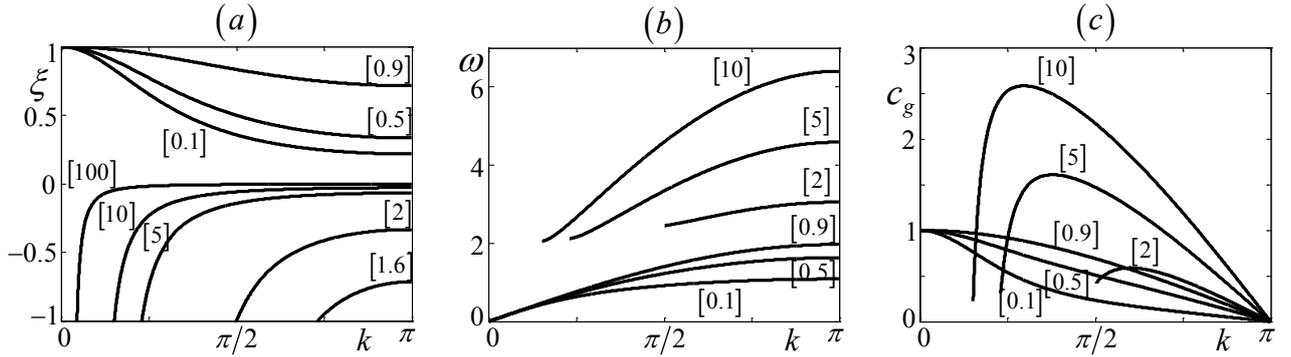

Fig. 14 Dispersion curves $\xi(k)$ - (a), $\omega(k)$ - (b) and $c_g(k)$ - (c) (the number in square brackets is $\gamma_m$).

In the case of in-phase oscillations ($\xi$ is positive), dispersion curves exist for all $k$ inside the BZ, while anti-phase oscillations are found only in BZ parts.

For a rather high $\gamma_m$, decay factors and group velocities are very sensitive to a small change of $\omega$ in the transition point vicinity. This fact suggests that the yield on the steady-state process will require relatively large time.

### 4.2. Transient processes. Interconnection of beaming and waveguide phenomena

Below, we present results of computer simulations of wave propagation processes in a stratified structure with a single layer of $n$-bonds – the case (a) of Subsection 4.1.1. The main goal of the analysis of transient solutions is to reveal a transition from the beaming wave pattern to the waveguide localization pattern. To excite the anti-symmetric oscillation form, we set the dipole source in $(0,0)$ and $(0,-1)$ nodes: $u_{0,0}(t) = \sin(\omega_0 t) H(t)$, $u_{0,-1}(t) = \sin(\omega_0 t) H(t)$.

In Fig. 15, snapshots of envelopes are shown at $t = 200$. Outside the envelope contours, perturbations remain less than 10% of the maximal amplitude in the source. Drawings (a) and (b) correspond to the stiffness $\gamma_n = 1$, and excitation frequencies are $\omega_0 = 2$ and $\omega_0 = 1+\sqrt{2}$, respectively. The former shows the resonant excitation of the diagonal-localized primitive waveform resulting in the beaming wave pattern [19], while in the latter – $\omega_0$ lies in the pass-band middle, $\left(2,\sqrt{8}\right)$ – such localization is significantly spread. Drawings (*c*), (*d*), (*e*), and (*f*) correspond to cases $\gamma = 1.05,\ 1.1,\ 1.5$ and $2.5$, respectively, for which decay factors are $\xi = -0.91,\ -0.83,\ -0.5$ and $-0.25$. For the each case, frequency $\omega_0$ is located in the middle of the corresponding pass-band:

$$\omega_0 = \frac{\omega(0)+\omega(\pi)}{2}.$$

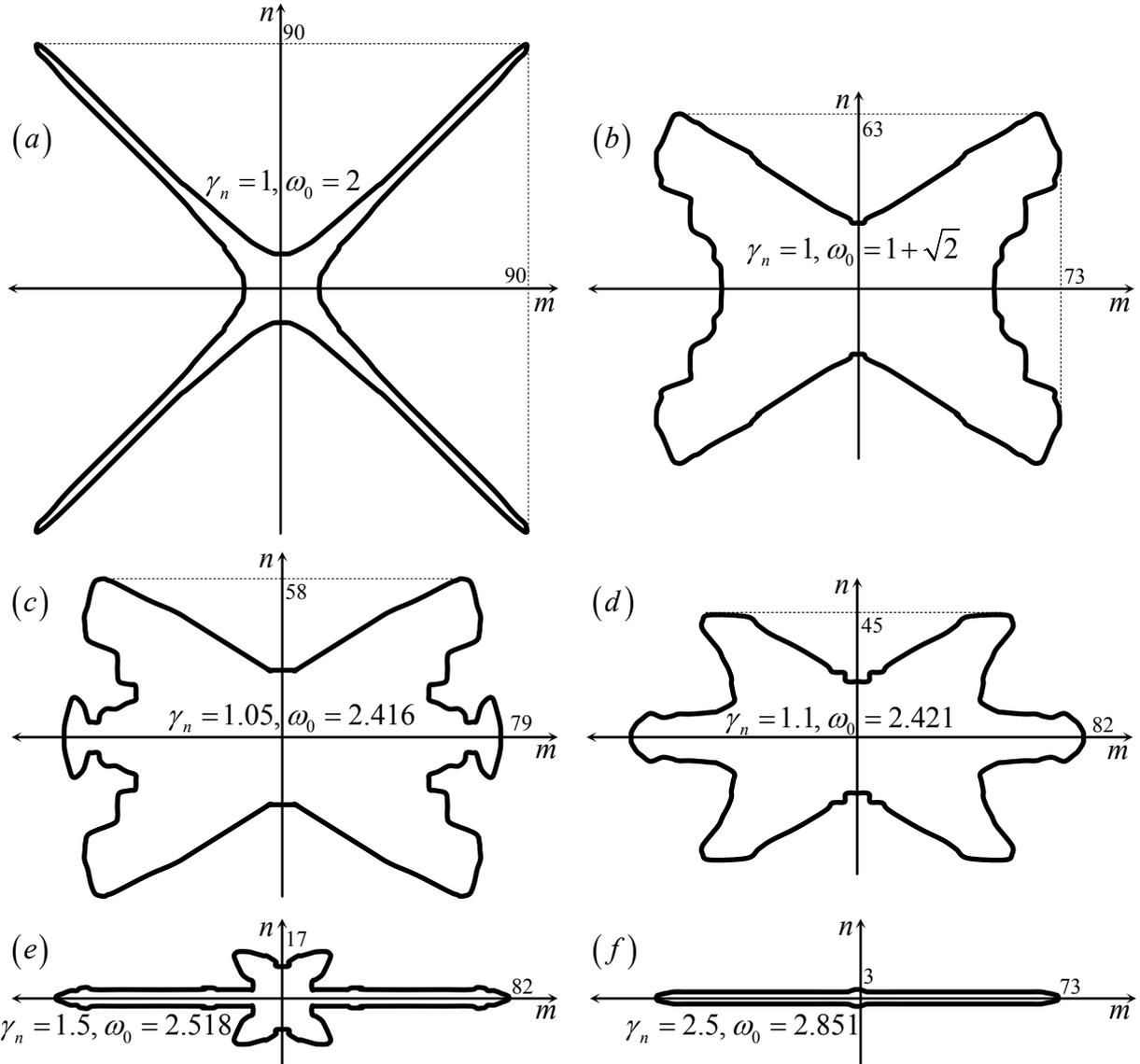

Fig. 15 The spatial distribution of disturbances at $t = 200$. Outside the contour, the displacements of nodes remain less than 10% of the maximal value in the source.

The analysis of distributions shows that with growing $\gamma_n$, the waveguide localization pattern becomes more distinct. The beaming-to-waveguide transition is observed up to the figure (*f*), where the former practically is not detected.

The above-mentioned transition can be seen more clearly in Fig. 16, where three snapshots at $t = 200, 400, 1000$ were calculated for excitation frequencies close to the lower limit of pass-band $\omega_0 = 1.05\omega_-$, $\omega_- = \sqrt{4\gamma_n^2/(2\gamma_n - 1)}$. Curves in columns (*a*) and (*b*) correspond to $\gamma_n = 1$ ($\omega_0 = 2.1$) and $\gamma_n = 1.1$ ($\omega_0 = 2.1087$), respectively.

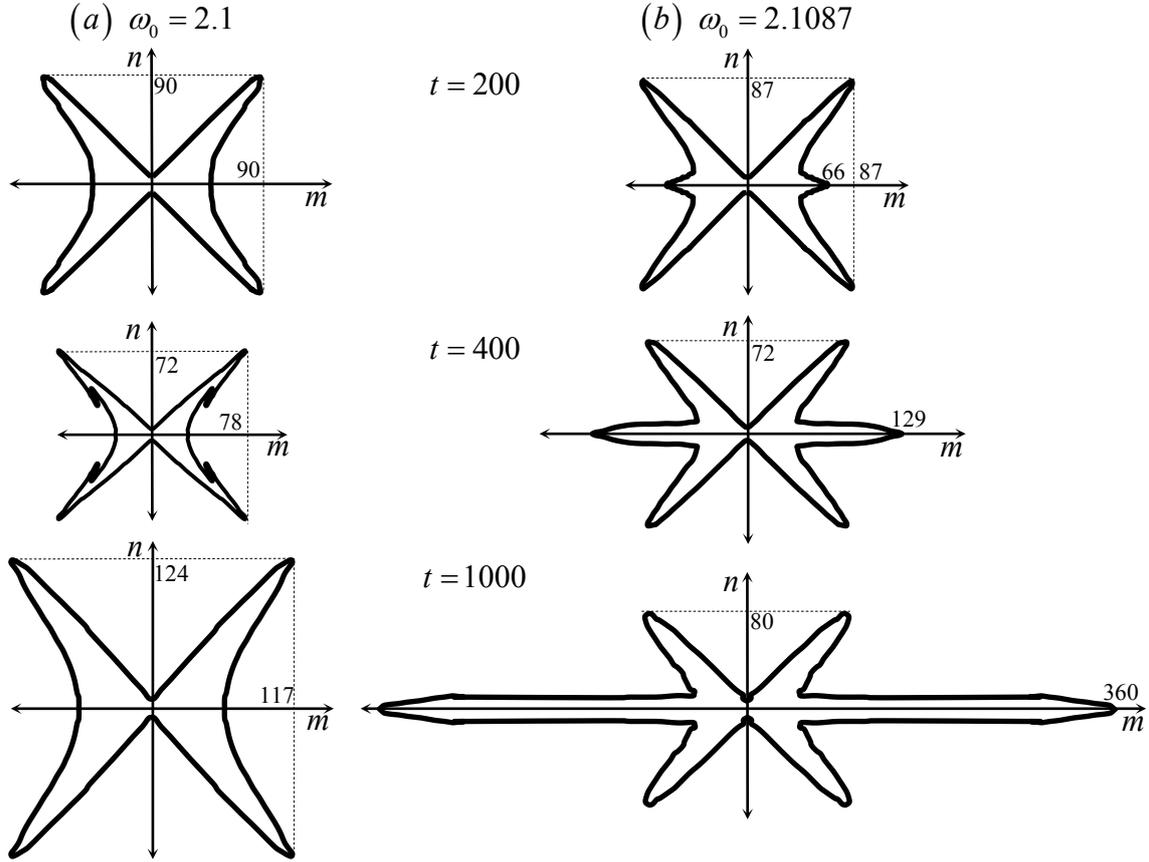

Fig. 16 Spatial distribution of disturbances versus time. Outside the contours, the displacements of nodes remain less than 10% of the maximal value in the source. (*a*) - $\omega_0 = 2.1$, (*b*) - $\omega_0 = 2.1087$.

Although frequency difference in the two comparable cases is relatively small (< 0.5%), the behavior of lattices becomes drastically different with time. In the non-waveguide case (*a*), $\omega_0$ differs by 10% from the resonant frequency $\omega = 2$ at which the diagonal beaming is realized [19], but in spite of this, the beaming pattern is observed for a relatively long time. The 'echo' of the beaming process holds long enough also in the waveguide case (*b*). As one can see, its contribution to the wave pattern is practically the same for all times considered, even when the waveguide localization effect is clearly dominating (see the contour at $t = 1000$).

### 4.3. Steady- and unsteady-state solutions for the source frequency inside the pass-band

In [31], an asymptotic solution to the steady-state problem 4.1.1 (*a*) is obtained in the case of the action of a dipole monochromatic source that consists of opposite forces applied to nodes $(m=0, n=0)$ and $(m=0, n=-1)$: $Q_{0,n} = (-1)^n \sin(\omega_0 t)$. Recall that the solution (19) with $|\xi|<1$ can exist only if $\gamma_n > 1$. If $\gamma_n < 1$ ($|\xi|>1$), then an exponentially growing (nonphysical) wave appears, while if $\gamma_n = 1$, then the wave represents anti-phase oscillations of constant amplitudes.

Equations of motion of the system in the steady-state formulation are:

$$\begin{cases} \ddot{u}_{m,n} = u_{m+1,n} + u_{m-1,n} + u_{m,n+1} + u_{m,n-1} - 4u_{m,n} & , n \neq 0,-1 \\ \ddot{u}_{m,0} = \gamma_m(u_{m-1,0} - 2u_{m,0} + u_{m+1,0}) + (u_{m,1} - u_{m,0}) + \gamma_n(u_{m,-1} - u_{m,0}) + \sin(\omega_0 t) & , n = 0 \\ \ddot{u}_{m,-1} = \gamma_m(u_{m-1,-1} - 2u_{m,-1} + u_{m+1,-1}) + (u_{m,-2} - u_{m,-1}) + \gamma_n(u_{m,0} - u_{m,-1}) - \sin(\omega_0 t) & , n = 0. \end{cases} \quad (31)$$

The solution of (31) is the following:

$$[u_{m,n}] = (-1)^n [U] \sin(\omega_0 t - k_0 m), \quad [U] = \frac{1-\xi^2}{2\sin k_0} \quad (n=0,-1; m \to \infty), \quad (32)$$

where square brackets are used to denote the steady-state amplitudes, and $k_0$ is the wave number corresponding to $\omega_0$ in the dispersion equation (22).

Below we compare this steady-state solution with the results of computer simulations of nonstationary problems described by the system (31) with zero initial conditions and non-stationary forces $Q_{0,n} = (-1)^n \sin(\omega_0 t) \cdot H(t)$ in the right-hand side. For a given frequency $\omega_0$, the comparable parameters are: amplitudes of displacements in nodes, $[U]$, (group) velocity of the propagating wave, $c_0$, and wavelength, $\lambda_0 = 2\pi/k_0$. The two latter parameters are obtained from dispersion relations.

Calculation results presented in Fig. 17 correspond to the case $\gamma_n = 2.5$ (in this case $\xi = -0.25$) for three values of $\omega_0$: $\omega_0 = 2.6$ (*a*), $2.8$ (*b*), and $3.1$ (*c*). These frequencies are located within the pass-band: $[\omega^-, \omega^+] = [2.50, 3.20]$. The chosen frequency, $\omega_0$, determines wave number, $k_0$ (and wavelength, $\lambda_0 = 2\pi/k_0$), which in its turn allows to find group velocity, $c_0$ (see Fig. 10). The dispersion parameters for the considered values of $\omega_0$ are given in Table 2. Curves in the left column of Fig. 17 numbered 1, 2 and 3 are envelopes (their positive parts) of nodes displacements in *m*axis $(m,n) = (0,0)$, $(25,0)$ and $(50,0)$ respectively, while the right column consists of $u_{m,0}$, $u_{m,1}$, $u_{m,2}$ distributions along $m$ axis taken at $t = 500$. The curves differ in their thicknesses: a thicker curve corresponds to a greater $m$ (the left column) or $n$ (the right column). The dashed perimeter is a moving (with the velocity $c_0$) steady-state solution $[U]$.

Table 2 Parameters of the stationary solution

| $\omega_0$ | $k_0$ | $\lambda_0$ | $c_0$ | $[U]$ |
|---|---|---|---|---|
| 2.6 | 0.7303 | 8.6040 | 0.2566 | 0.7027 |
| 2.8 | 1.3643 | 4.6053 | 0.3496 | 0.4789 |
| 3.1 | 2.3186 | 2.7100 | 0.2365 | 0.6393 |

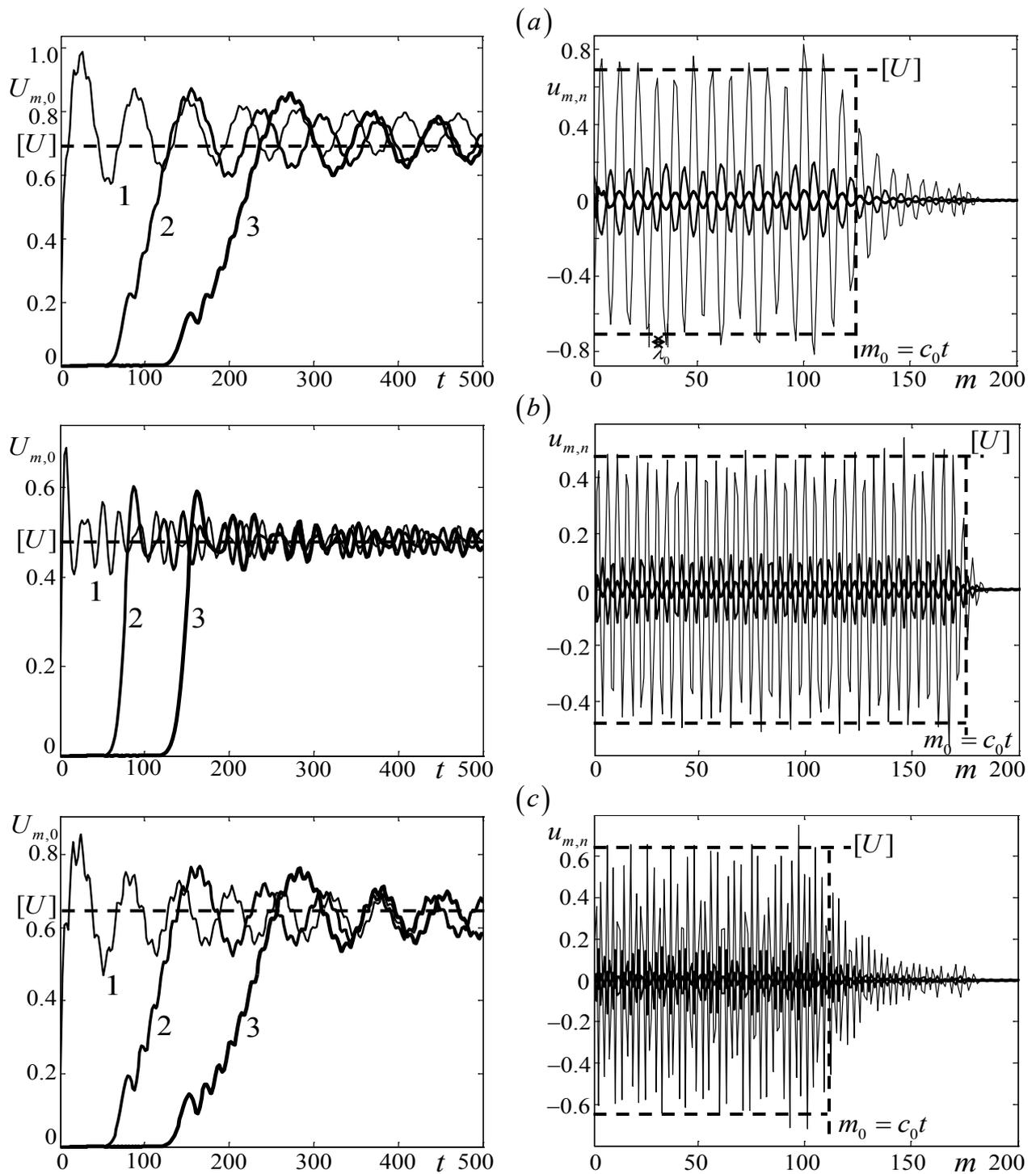

Fig. 17 Comparison of transient solutions with the steady-state asymptotes. (*a*) $\omega_0 = 2.6$, (*b*) $\omega = 2.8$, (*c*) $\omega_0 = 3.1$.

The results shown in the left column allow numerical estimation of steady-state magnitudes as average amplitudes, $[U]$, of calculated envelopes. These estimations practically immediately coincide with those predicted by the solution (32) after the arrival of major perturbations to the indicated node. As one can see, these perturbations move at the velocity equal to $c_0$ – the group velocity determined by dispersion relations. Results in the right column show that relative average magnitudes of $u_{m,n}$ are in good accordance with the decay factor $\xi = -0.25$ (which results in $u_{m,n} = (-0.25)^n u_{m,0}$).

The data in the right column serve to determine wavelength $\lambda_0 = 2\pi / k_0$, to detect the peculiarities of the transient process and the correspondence of both solutions. Dashed straight lines are the steady-state amplitude, $[U]$, taken from analytical asymptote (32). One can see that a wavy character of calculated envelopes has a tendency to spread with time; the speed of the stabilization is maximal for the frequency $\omega_0$ located in the median part of the pass-band. At the same time, the steady-state amplitudes can also be satisfactorily estimated in two other cases, where frequencies are relatively close to pass/stop band interfaces. The transient waveform with respect to the transversal axis $n$ is satisfactorily determined by the steady-state estimate: $u_{m,n} = \xi^n u_{m,0}$.

### 4.4. Unsteady-state solutions at the interfaces of pass- and stop-bands. Resonant waves

Below, together with the waveguide localization phenomenon, we consider the formation of resonant waves in the case where the excitation frequency coincides with the demarcating pass- and stop-bands. The used technique is based on the method of reversion of integral Fourier-Laplace transforms in nonstationary problems for continuous structures [4] and its extension to the case of lattice structures [17, 18].

We have obtained an analytical solution for the short-wave resonance on the basis of the simplest MSC, whose dispersion curve in the vicinity of $k = \pi$ is similar to that in the considered case of a stratified lattice, as was shown above in Subsection 4.1.1. The corresponding Cauchy problem is formulated as follows ($m$ is the node number, $M = g = 1$ are measurement units)

$$\begin{cases} \ddot{u}_m = u_{m+1} - 2u_m + u_{m-1} & , m \neq 0 \\ \ddot{u}_0 = u_1 - 2u_0 + u_{-1} + Q(t) & , m = 0 \\ u_n(0) = \dot{u}_n(0) = 0 \end{cases} \quad (33)$$

where $Q(t)$ is the outside force applied to 0-node. In the MSC case, the resonant frequency demarcating pass- and stop-bands at the end $k = \pi$ of the BZ is $\omega_0 = \omega(\pi) = 2$. Let the external monochromatic force be $Q(t) = 2e^{i\omega_0 t} H(t)$, $\omega_0 = 2$.

We apply to the system (33) the discrete Fourier transform with respect to $m$ (parameter $k$), $u^{F_d}(k,t) = \sum_{m=-\infty}^{m=\infty} u_m(t) e^{ikm}$, and the Laplace transform $u^{LF_d}(k,t) = \int_0^\infty u^{F_d}(k,t) e^{-pt}$ with respect to time (parameter $p$). The formal solution of (33) in the originals can be written as

$$u_m(t) = \frac{1}{2\pi^2 i} \int_{-\pi}^{\pi} \int_{i\sigma-\infty}^{i\sigma+\infty} \frac{e^{pt-ikm}}{(p-\omega_0 i)(p^2 + 4\sin^2(k/2))} dp\, dk. \quad (34)$$

After the inversion of the Laplace transform, we obtain

$$u_m(t) = \frac{e^{\omega_0 it}}{\pi} \int_0^\pi \frac{\left[1 - e^{-\omega_0 i(1-\sin(k/2))t}\right]}{1-\sin(k/2)} e^{-ik|m|} dk. \quad (35)$$

The integral function in (35) is limited, it possesses a removable singularity at $k = \pi$ and, due to a general feature of trigonometric integrals, tends to zero at $m \to \infty$ for any finite $t$. Omitting technical details related to obtaining the asymptotic ($t \to \infty$) expression of integral (35), we have obtained the following final asymptotic representation of the solution:

$$u_n(t) \sim -\frac{i\sqrt{t} e^{i(\omega_0 t - \pi m)}}{\pi} \left[F_1(\eta) + iF_2(\eta)\right], \quad \eta = 2|m|/\sqrt{t},\ t \to \infty. \quad (36)$$

Functions $F_1(\eta)$ and $F_2(\eta)$ are well-known in the theory of unsteady state waves (see [4]) and have the following form:

$$F_1(\eta) = \int_0^\infty \frac{\cos(\eta z)\sin(z^2)}{z^2} dz = \sqrt{\frac{\pi}{2}}\left(\cos\frac{\eta^2}{4} + \sin\frac{\eta^2}{4}\right) - \frac{\pi|\eta|}{2}\left[C\left(\frac{\eta^2}{4}\right) - S\left(\frac{\eta^2}{4}\right)\right],$$

$$F_2(\eta) = \int_0^\infty \frac{\cos(\eta z)(1-\cos(z^2))}{z^2} dz = \sqrt{\frac{\pi}{2}}\left(\cos\frac{\eta^2}{4} - \sin\frac{\eta^2}{4}\right) - \frac{\pi|\eta|}{2}\left[1 - C\left(\frac{\eta^2}{4}\right) - S\left(\frac{\eta^2}{4}\right)\right], \quad (37)$$

where $C(z)$ and $S(z)$ are Fresnel integrals: $C(z) = \frac{1}{\sqrt{2\pi}} \int_0^z \frac{\cos\tau}{\sqrt{\tau}} d\tau$, $S(z) = \frac{1}{\sqrt{2\pi}} \int_0^z \frac{\sin\tau}{\sqrt{\tau}} d\tau$.

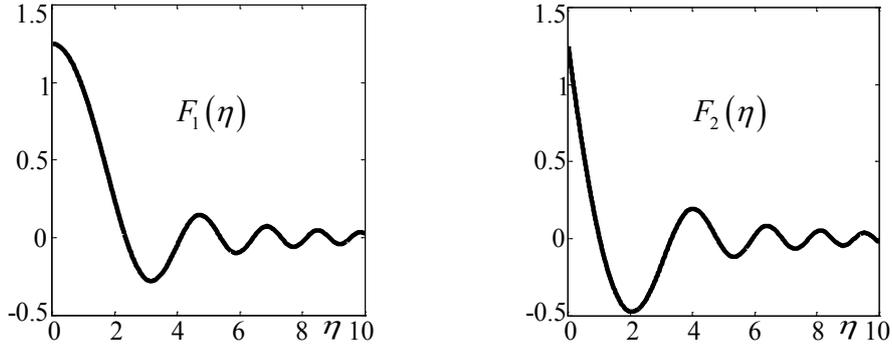

Fig. 18 Functions $F_1(\eta)$ - (*a*) and $F_2(\eta)$ - (*b*).

Functions $F_1(\eta)$ and $F_2(\eta)$ are finite for any $m$ and tend to zero at $\eta \to \infty$ (Fig. 18). Real and imaginary parts of the solution (36) corresponding to the cosine source $Q = 2\cos(\omega_0 t) H(t)$ and sine source $Q = 2\sin(\omega_0 t) H(t)$, respectively, are written as

$$\mathrm{Re}[U_m(t)] \sim \frac{\sqrt{t}}{\pi}\left[F_2(\eta)\cos(\omega_0 t - \pi m) + F_1(\eta)\sin(\omega_0 t - \pi m)\right],$$

$$\mathrm{Im}[U_m(t)] \sim \frac{\sqrt{t}}{\pi}\left[F_2(\eta)\sin(\omega_0 t - \pi m) - F_1(\eta)\cos(\omega_0 t - \pi m)\right]. \tag{38}$$

The functions $F_1(\eta)$ and $F_2(\eta)$ represent envelopes of oscillations with a carrier frequency $\omega = \omega_0$. Recall that in the case of the resonant excitation at frequencies in the BZ ends $k = 0$ and $\pi$, where the group velocity $c_g = 0$, the velocity of wave propagation depends on the shape of the dispersion curve in the vicinities of $k = 0$ and $\pi$ (more precisely, on the order of the first nonzero derivative $\partial^n \omega / \partial k^n$ at these points [17]). As shown above, in Subsection 4.1.1, dispersion curves for the considered stratified lattices can be precisely approximated by those for the MSC with specially chosen parameters. Obtained asymptotic formulas for the shortwave resonance at $k = \pi$, $\omega_0 = \omega(\pi) \equiv \omega^+$ can be generalized for the case of the longwave resonance at $k = 0$, $\omega_0 = \omega(0) \equiv \omega^-$. The described resonance wave moves along the waveguide with deceleration proportional to $\sqrt{t}$ – see expression for $\eta$ in (36) – similar to the heat propagation law. Within this process, amplitudes $u_n(t)$ increase with time as $\sqrt{t}$.

Below we present examples of numerical simulations of the resonant excitations. We also compare resonant waves and near-resonant waves possessing a steady state form. Due to the fact that in these cases the nonstationary steady-state transition requires a relatively long time, the mentioned comparison allows elucidating in detail the peculiarities of the resonance development. In Fig. 19, the results of the numerical simulation are presented related to the first resonance, (*a*), $\omega_0 = \omega^-$, to the second one, (*d*), $\omega_0 = \omega^+$, and for two near-resonant cases: (*b*) $\omega_0 = 1.005\omega^-$, and (*c*) $\omega_0 = 0.995\omega^+$. All calculations are conducted for the case $\gamma = 2.5$ (decay factor $\xi = -0.25$). Dispersion parameters and the steady-state solution $[U]$ for the above-considered values of $\omega_0$ are presented in Table 3.

Table 3 Parameters of the stationary solution

| $\omega_0$ | $k_0$ | $\lambda_0$ | $c_0$ | $[U]$ |
|---|---|---|---|---|
| 2.5000 | 0 | $\infty$ | 0 | – |
| 2.5125 | 0.2510 | 25.0356 | 0.0988 | 1.8875 |
| 3.1856 | 2.8208 | 2.2277 | 0.0991 | 1.4851 |
| 3.2016 | $\pi$ | 2.0000 | 0 | – |

As in the previous Fig. 17, curves in the left column show positive parts of envelopes versus time in the nodes indicated inside the figure, dashed straight lines correspond to steady-state solution (32). Distributions of displacements $u_{m,0}$, $u_{m,1}$ and $u_{m,2}$ along the *m* axis at fixed values of time are depicted in the right columns. One can observe a contra-phase oscillation form realized along the *n*-axis of resonant and near-resonant processes, which practically coincide on the time interval where the latter has not peaked yet (see the left column). After that, envelopes of the near-resonant processes acquire an oscillation character, while resonant ones monotonically increase as $\sqrt{t}$, with the same growth that was analytically estimated for the case of a MSC waveguide. Thus, computer simulations confirm the applicability of the asymptotic solution obtained on the basis of the simplest model.

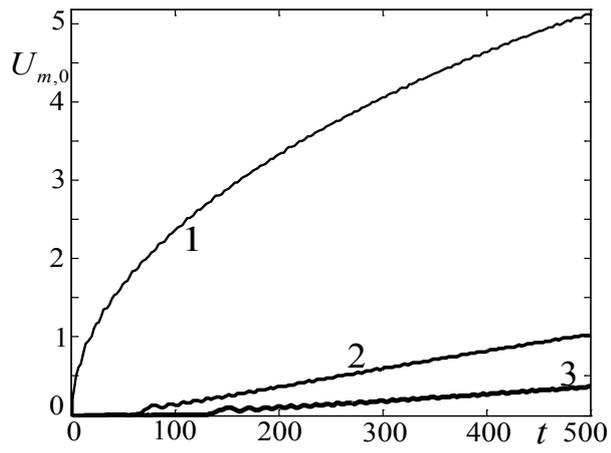
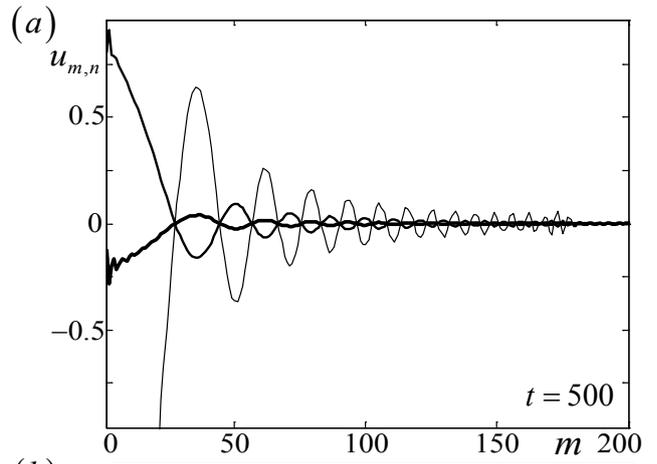
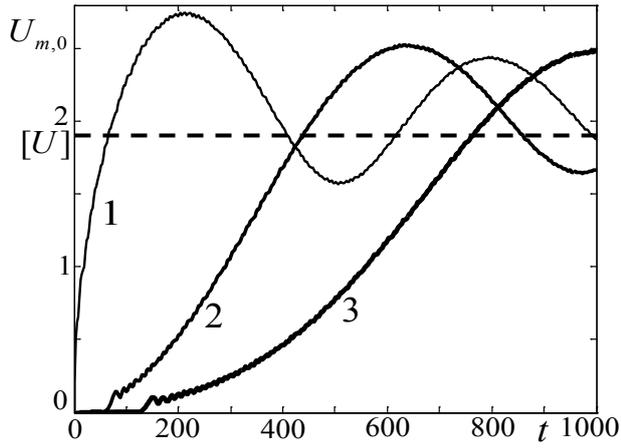
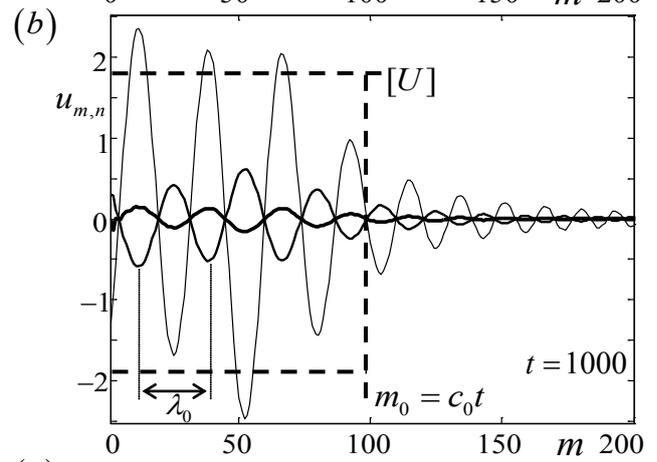
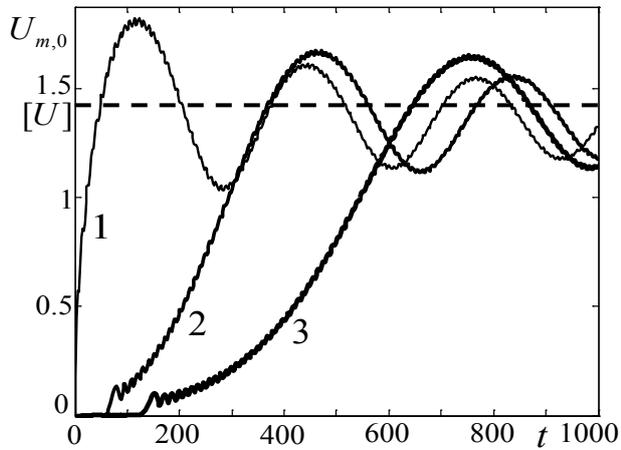
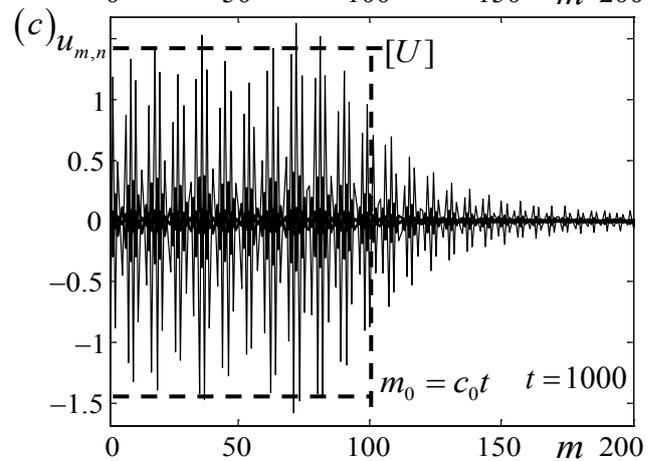
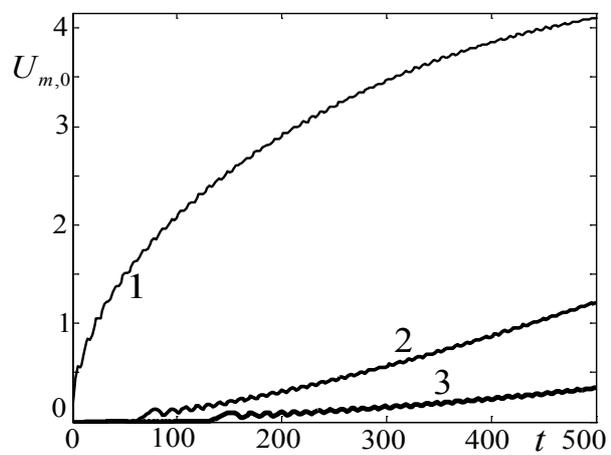
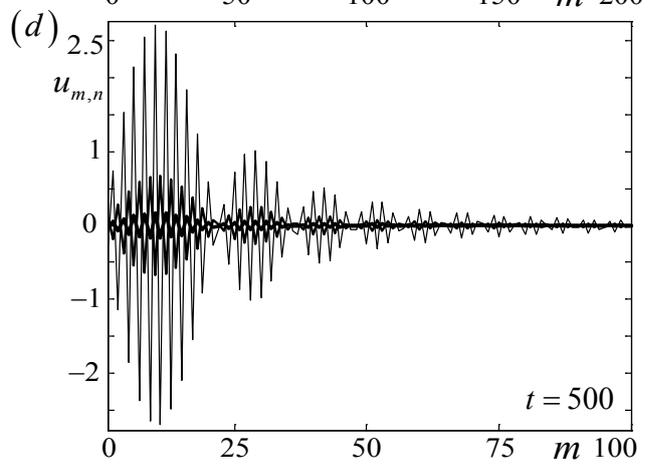

Fig. 19 Resonant and near-resonant patterns in a stratified square-cell lattice ($\gamma_n = 2.5$): (a) - $\omega_0 = \omega^-$, (b) - $\omega_0 = 1.005\omega^-$, (c) - $\omega_0 = 0.995\omega^+$ and (d) - $\omega_0 = \omega^+$.

## 5. Embedded three-chain stratum with different stiffnesses

### 5.1. Dispersion analysis

We consider a stratified structure consisting of two lattices linked by a layer as shown in Fig. 20. System parameters are: $M_1 = M_2 = 1$, $g_{1,x} = g_{1,y} = g_{2,x} = 1$, $\gamma_n \equiv g_{2,y} \neq 1$. As in the previous sections, we aim to reveal the frequency spectra in which the embedded stratum can play the role of a peculiar trap for energy emitted by a monochromatic source.

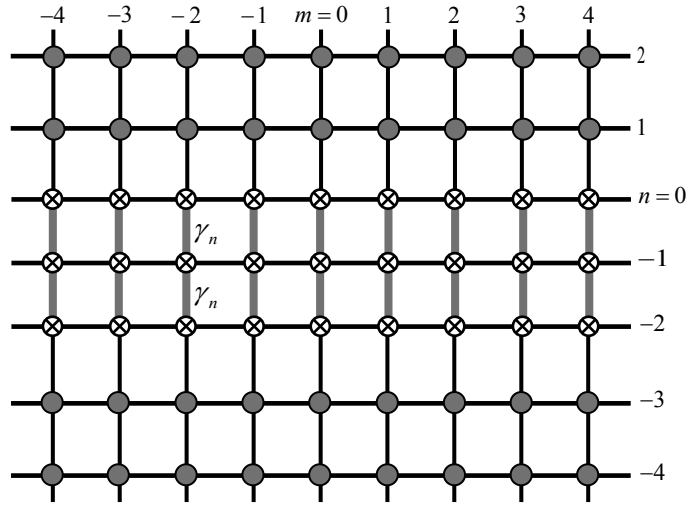

Fig. 20 Embedded 3-chain stratum.

In this case, the system (1) is transformed into the following:

$$\begin{cases} \ddot{u}_{m,n} = u_{m+1,n} + u_{m-1,n} + u_{m,n+1} + u_{m,n-1} - 4u_{m,n} & , n > 0 \\ \ddot{u}_{m,0} = (u_{m-1,0} - 2u_{m,0} + u_{m+1,0}) + \gamma_n(u_{m,-1} - u_{m,0}) + (u_{m,1} - u_{m,0}) & , n = 0 \\ \ddot{u}_{m,-1} = (u_{m+1,-1} - 2u_{m,-1} + u_{m-1,-1}) + \gamma_n(u_{m,0} - 2u_{m,-1} + u_{m,-2}) & , n = -1 \\ \ddot{u}_{m,-2} = (u_{m-1,-2} - 2u_{m,-2} + u_{m+1,-2}) + \gamma_n(u_{m,-1} - u_{m,-2}) + (u_{m,-3} - u_{m,-2}) & , n = -2 \\ \ddot{u}_{m,n} = u_{m+1,n} + u_{m-1,n} + u_{m,n+1} + u_{m,n-1} - 4u_{m,n} & , n < -2. \end{cases} \quad (39)$$

To reveal the existence of the stratum waveguide, we seek a solution to system (39) in the form of a wave moving in the $m$-direction with the magnitude vanishing at $n \to \pm\infty$ (the structure is symmetric with respect to the $m$ axis; consequently, decay factors in the parts $n > 0$ and $n < 2$ are the same):

$$\begin{cases} u_{m,n}(t) = \xi^n \cdot e^{i(\omega t \pm km)} & , n > 0 \\ u_{m,n}(t) = A_n \cdot e^{i(\omega t \pm km)} & , -2 \leq n \leq 0 \\ u_{m,n}(t) = \xi^{-(n+2)} \cdot e^{i(\omega t \pm km)} & , n < -2. \end{cases} \quad (40)$$

After substituting (40) into (39) we obtain $A_0 = A_{-2} = 1$ and an algebraic system to determine $A_{-1}(k), \xi(k)$ and $\omega(k)$:

$$\begin{cases} \omega^2 = 4\sin^2(k/2) + 2 - \xi - \xi^{-1} \\ \omega^2 = 4\sin^2(k/2) + \gamma_n(1 - A_{-1}) + 1 - \xi \\ A_{-1}\omega^2 = 4A_{-1}\sin^2(k/2) - 2\gamma_n(1 - A_{-1}). \end{cases} \quad (41)$$

The dispersion relation of the considered system is $\omega = \sqrt{2 + \Psi - \Gamma}$, while $\xi$ and $A_{-1}$ are found from the following equations:

$$A_{-1} = \frac{\gamma_n - 1 + \xi^{-1}}{\gamma_n}, \qquad \xi = \frac{\gamma_n(1 - A_{-1})(A_{-1} + 2) + A_{-1}}{A_{-1}}. \quad (42)$$

Solving (42), taking into account the requirement $|\xi| < 1$, we obtain:

$$A_{-1} = 1 - \frac{1}{\gamma_n} + \frac{2\gamma_n - 2}{3\gamma_n^2 - 2\gamma_n - \gamma_n\sqrt{9\gamma_n^2 - 8\gamma_n}}, \quad \xi = \frac{3\gamma_n - 2 - \sqrt{9\gamma_n^2 - 8\gamma_n}}{2\gamma_n - 2}, \quad \gamma_n > 1. \quad (43)$$

Both $\xi$ and $A_{-1}$ depend on $\gamma_n$ and do not depend on $k$, as in the case for a single stratum, 4.1.1(*a*) (dispersion curves exist in the entire BZ spectrum).

Decay and amplitude factors as functions of $\gamma_n$ are presented in Fig. 21 (*a*) while Fig. 21 (*b*) and (*c*) refer to frequency, $\omega(k)$, and group velocity, $c_g(k)$, depicted for $\gamma_n = 1, 1.5, 2, 3$ and 5. As in Fig. 10, the limiting frequencies and group velocities corresponding to $\gamma_n = 1$ are shown by dashed curves.

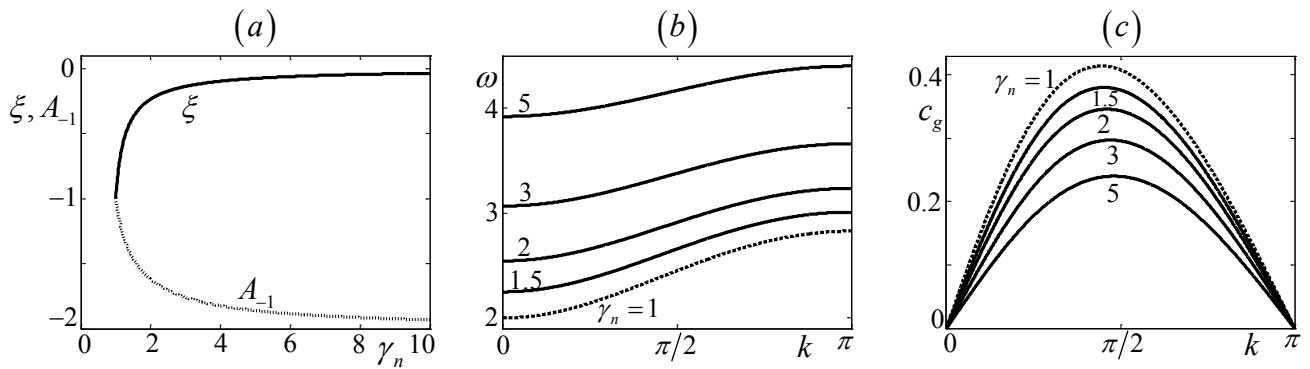

Fig. 21 Decay and amplitude factors, (*a*), frequency, (*b*), and group velocity, (*c*).

## 6. Conclusions

The phenomenon of propagating wave localization along *an inner layer* in a square-cell sandwiched lattice is discovered. Steady-state and nonstationary antiplane elastic problems are solved. The latter corresponds to point monochromatic excitations with frequencies within the pass-band. Parameters of localization, namely, the pass-band where *the* inner lattice *layer* traps the main part of the source energy, energy flux velocity and decay factors determining wave attenuation along the direction normal to *the layer* are analyzed depending on the *layer* properties. A semi-infinite lattice covered by *a layer* is considered as a special case. A similarity of the considered 2D localization process and the 1D waveguide propagation in a mass-spring chain of reduced parameters is found.

An unexpected behavior of the roots of the dispersion equation is observed depending on *layer* parameters – the localization holds only for a certain part of wavelengths inside the Brillouin zone. Frequencies corresponding to the *limiting wavelengths* are called "*transition points*".

Steady-state and nonstationary solutions are compared in order to reveal the contribution of the latter into the localization process. In the cases where the nonstationary solution has a stationary limit, a good correspondence between these solutions is obtained. It is shown how the nonstationary contribution increases when approaching the ends of the Brillouin zone and transition points.

The most important results include the analysis of particularities of (*a*) transient waves in cases where steady-state solutions are absent – resonant waves are discovered, whose frequencies demarcate pass- and stop-bands (at the ends of the Brillouin zone), and (*b*) wave processes in the vicinities of transition points. In the former case, it is shown that perturbations in resonant waves increase as $\sqrt{t}$ – the same growth as that estimated for the 1D mass-spring chain. Thus, computer simulations confirm the applicability of the obtained asymptotic solutions on the basis of the simplest model. In the latter case, a simultaneous onset of two different localization phenomena is obtained – a spatial star-like beaming and one-dimensional waveguide-like localization.

**Acknowledgments**

We gratefully acknowledge Professor L.I. Slepyan for valuable contribution to the work and insightful discussions. This research was supported by the Israel Science Foundation, Grant 504/08.

**References**

[1] Lord Rayleigh, On the maintenance of vibrations by forces of double frequency, and on the propagation of waves through a medium endowed with a periodic structure, Philosophical Magazine 24 (1887) 145-159.

[2] L. Brillouin, Wave Propagation in Periodic Structures, Dover Publications Inc, New York, 1953.

[3] C.T. Sun, J.D. Achenbach, G. Herrmann, Time-harmonic waves in a stratified medium propagating in the direction of the layering, *Journal of Applied Mechanics* 35 (1968) 408- 411.

[4] L. I. Slepyan, Non-steady-state elastic waves, Sudostroenie, Leningrad, 1972 (in Russian).

[5] D.J. Mead, A general theory of harmonic wave propagation in linear periodic systems with multiple coupling, *Journal of Sound and Vibration* 27 (1973) 235-249.

[6] J.D. Achenbach, Waves and vibrations in directionally reinforced solids, *Composite Materials* 2 (1974), Academic Press, 310-352.

[7] T.J. Delph, G. Herrmann, R.K. Kaul, Harmonic wave propagation in a periodically layered, infinite elastic body, antiplane strain, *Journal of Applied Mechanics* 45 (1978) 343-349.

[8] A.S. Mal, Wave propagation in layered composite laminates under periodic surface loads, *Wave Motion* 10 (1988) 257-266.

[9] D.J. Mead, Y. Yaman, The response of infinite periodic beams to point harmonic forces. A flexural wave analysis, *Journal of Sound and Vibration* 144 (1991) 507-522.

[10] D.J. Mead, R.G. White, X.M. Zhang, Power transmission in a periodically supported infinite beam excited at a single point, *Journal of Sound and Vibration* 169 (1994) 558-561.

[11] D.J. Mead, Wave propagation in continuous periodic structures: research contributions from Southampton, 1964-1995, *Journal of Sound and Vibration* 190 (1996) 495-524.

[12] M.M. Sigalas, E.N. Economou, Elastic and acoustic wave band structure, *Journal of Sound and Vibration* 158 (1992) 377-389.

[13] E. Yablonovitch, Photonic band-gap crystals, *Journal of Physics: Condensed Matter* 5 (1993) 2443-2460.

[14] M.S. Kushwaha, P. Halevi, L. Dobrzynski, B. Djafari-Rouhani, Acoustic band structure of periodic elastic composites, *Physical Review Letters* 71 (1993) 2022-2025.

[15] J.S. Jensen, Phononic band gaps and vibrations in one- and two-dimensional mass-spring structures, *Journal of Sound and Vibration* 226 (2003) 1053-1078.

[16] P. G. Martinsson, A. B. Movchan, Vibrations of lattice structures and phononic band gaps, *The Quarterly Journal of Mechanics and Applied Mathematics* 56 (2003) 45–64.

[17] L.I. Slepyan, O.V. Tsareva, Energy flux for zero group velocity of the current wave, *Soviet Physics - Doklady* 32 (1987) 522-526.

[18] G. Osharovich, M. Ayzenberg-Stepanenko, O. Tsareva, Wave propagation in elastic lattices subjected to a local harmonic loading. I. A quasi-one-dimensional problem, *Journal of Continuum Mechanics and Thermodynamics* 22 (2010) 581-597.

[19] M. Ayzenberg-Stepanenko, L. Slepyan, Resonant-frequency primitive waveforms and star waves in lattices, *Journal of Sound and Vibration* 313 (2008) 812-821.

[20] G. Osharovich, M. Ayzenberg-Stepanenko, O. Tsareva, Wave propagation in elastic lattices subjected to a local harmonic loading. II. Two-dimensional problems, *Journal of Continuum Mechanics and Thermodynamics* 22 (2010) 599-616.

[21] L.I. Slepyan, M.V. Ayzenberg-Stepanenko, Localized transition waves in bistable-bond lattices, *Journal of the Mechanics and Physics of Solids* 52 (2004) 1447-1479.

[22] G.S. Mishuris, A.B. Movchan, L.I. Slepyan, Localised knife waves in a structured interface, *Journal of the Mechanics and Physics of Solids* 57 (2009) 1958-1979.

[23] L.I. Slepyan, G.S. Mishuris, A.B. Movchan, Crack in a lattice waveguide, *International Journal of Fracture* 162 (2010) 91–106.

[24] M. Ayzenberg-Stepanenko, E. Sher, G. Osharovich, Waveguides in periodically structured solids, *Proceedings of the 8th Israeli-Russian Bi-National Workshop*, Jerusalem, 2009, pp. 161-174 (http://www.ariel.ac.il/management/research/pf/zinigrad/mmt/WS2009/Papers/161-174.doc).

[25] G. Osharovich, M. Ayzenberg-Stepanenko, E. Sher, Unexpected wave-oscillation effects in lattices of regular structure, *Proceedings of the 8th Israeli-Russian Bi-National Workshop*, Jerusalem, 2009, pp. 59-73

(http://www.ariel.ac.il/management/research/pf/zinigrad/mmt/WS2009/Papers/59-73.doc).

[26] V.A. Saraikin, M.V. Stepanenko, O.V. Tsareva, Elastic waves in a medium of block structure, *Journal of Mining Science* 25 (1988) 14-23.

[27] R.S. Langley, The response of two-dimensional periodic structures to point harmonic forcing, *Journal of Sound and Vibrations* 197 (1996) 447-469.

[28] R.S. Langley, The response of two-dimensional periodic structures to impulsive point loading. *Journal of Sound and Vibrations* 201 (1997) 235–253.

[29] R.S. Langley, N.S. Bardell, H.M. Ruivo, The response of two-dimensional periodic structures to harmonic point loading: a theoretical and experimental study of a beam grillage, *Journal of Sound and Vibrations* 207 (1997) 521–535.

[30] G.S. Mishuris, A.B. Movchan, L.I. Slepyan, Waves and fracture in an inhomogeneous lattice structure, *Waves in Random and Complex Media* 17 (2007) 409-428.

[31] L.I. Slepyan, Personal communication, 2009.